\documentclass[twocolumn]{aastex61}  
\usepackage{amsmath}

\def\HII{{\ion{H}{2}}}
\def\OII{[{\ion{O}{2}}]}
\def\4959_5007{[\ion{O}{3}]~$\lambda \lambda$4959,5007}
\def\OIII49595007{[\ion{O}{3}]~$\lambda \lambda 4959,5007$}
\def\ratioR23{([\ion{O}{2}]~$\lambda \lambda$3727,9 + [\ion{O}{3}]~$\lambda\lambda$4959,5007)/H$\beta$}
\def\R23{${\rm R}_{23}$}
\def\dS23{${\rm S}_{23}$}

\def\lOIII{[\ion{O}{3}]~$\lambda$5007}
\def\lNII{[{\ion{N}{2}}]~$\lambda$6584}

\def\NII{[{\ion{N}{2}}]}

\def\ratioS23{([\ion{S}{2}]~$\lambda \lambda$6717,31 +[\ion{S}{3}]~$\lambda\lambda$9069,9532)/H$\beta$}
\def\NIIHa{[\ion{N}{2}]/H$\alpha$}
\def\SIIHa{[\ion{S}{2}]/H$\alpha$}

\def\SII{[{\ion{S}{2}}]}

\def\Hb{{H$\beta$}}
\def\O4363{[{\ion{O}{3}}]~$\lambda$4363}
\def\OIII{[{\ion{O}{3}}]}

\def\Ha{{H$\alpha$}}
\def\Ep{$E_{\rm peak}$}

\newenvironment{DIFnomarkup}{}{}  

\accepted{March 1, 2018}

\shorttitle{Interrogating Seyferts with NebulaBayes}
\shortauthors{A. D. Thomas et al.}

\begin{document}

\title{Interrogating Seyferts with NebulaBayes: Spatially probing the narrow-line region radiation fields and chemical abundances}

\correspondingauthor{Adam D. Thomas}
\email{adam.thomas@anu.edu.au}

\author[0000-0003-1761-6533]{Adam D. Thomas}
\author{Michael A. Dopita}
\author{Lisa J. Kewley}
\author{Brent A. Groves}
\affiliation{RSAA, Australian National University, Cotter Road, Weston Creek, ACT 2611, Australia}
\affiliation{ARC Centre of Excellence for All Sky Astrophysics in 3 Dimensions (ASTRO 3D)}

\author{Ralph S. Sutherland}
\affiliation{RSAA, Australian National University, Cotter Road, Weston Creek, ACT 2611, Australia}

\author{Andrew M. Hopkins}
\affiliation{ARC Centre of Excellence for All Sky Astrophysics in 3 Dimensions (ASTRO 3D)}
\affiliation{Australian Astronomical Observatory, PO Box 915, North Ryde, NSW 1670, Australia}

\author{Guillermo A. Blanc}
\affiliation{Observatories of the Carnegie Institution for Science, 813 Santa Barbara Street, Pasadena, CA 91101, USA}

\begin{abstract}
NebulaBayes is a new Bayesian code that implements a general method of comparing observed emission-line fluxes to photoionization model grids.  The code enables us to extract robust, spatially resolved measurements of abundances in the extended narrow line regions (ENLRs) produced by Active Galactic Nuclei (AGN).  We observe near-constant ionization parameters but steeply radially-declining pressures, which together imply that radiation pressure regulates the ENLR density structure on large scales.  Our sample includes four `pure Seyfert' galaxies from the S7 survey that have extensive ENLRs.  NGC\,2992 shows steep metallicity gradients from the nucleus into the ionization cones.  An {\it inverse} metallicity gradient is observed in ESO\,138-G01, which we attribute to a recent gas inflow or minor merger.  A uniformly high metallicity and hard ionizing continuum are inferred across the ENLR of Mrk\,573.  Our analysis of IC\,5063 is likely affected by contamination from shock excitation, which appears to soften the inferred ionizing spectrum.  The peak of the ionizing continuum $E_{\rm peak}$ is determined by the nuclear spectrum and the absorbing column between the nucleus and the ionized nebula.  We cannot separate variation in this intrinsic $E_{\rm peak}$ from the effects of shock or \HII\ region contamination, but $E_{\rm peak}$ measurements nevertheless give insights into ENLR excitation.  We demonstrate the general applicability of NebulaBayes by analyzing a nuclear spectrum from the non-active galaxy NGC\,4691 using a \HII\ region grid.  The NLR and \HII\ region model grids are provided with NebulaBayes for use by the astronomical community.
\end{abstract}


\section{Introduction} \label{sec:Intro}

The impact of active galactic nuclei (AGN) on the evolution of their host galaxies has received considerable attention in the literature over the previous two to three decades \citep[for reviews see e.g.][]{KormendyHo2013_AGN_review, KingPounds_2015_AGN_feeback_review, Wagner_2016_AGN_feedback_review}.  However, many questions remain regarding the impacts of the `radiative' (or `quasar') mode and the `radio' mode of nuclear activity.  Challenges lie firstly in understanding the physical processes by which the radiative and mechanical energy released by accretion couples to the interstellar medium (ISM), and secondly in connecting the resulting outflows, extended narrow line regions (ENLRs), expanding radio lobes, and related phenomena to the concrete mechanisms by which they influence the ISM and hence star formation in the host.  

Advancing our understanding of these phenomena requires measuring the physical properties of the ISM.  A variety of integral field spectrographs are now routinely used to study active nuclei with high spatial and spectral resolution \citep[e.g.][]{StorchiBergmann_2010_N4151_NIFS, 2011_MullerSnachez_AGN_outflows, Cresci_2015_MUSE_AGN, Smajic_2015_N1566_SINFONI, Dopita_2015_S7_II}, but we currently lack the tools to fully interpret how the AGN influences the surrounding ISM.

For example, the determination of gas-phase abundances in AGN has received comparably less attention than abundance measurements in \HII\ regions.  It is often broad line region (BLR) abundances which are investigated when AGN abundances have been considered \citep[quasar BLR abundances have been studied for decades, e.g.][]{Davidson_Netzer_1979_AGN}.  Work by \citet{Dietrich_2003_BLR_Z}, \citet{Nagao_2006_SDSS_BLR_Z}, \citet{Juarez_2009_QSO_Z} and others on the nebular metallicity of quasar BLRs has shown that there is little evolution of BLR metallicity as a function of redshift, with generally high metallicities observed even for quasars at $z \sim 6$.  \citet{Nagao_2006_SDSS_BLR_Z} find that more luminous quasars tend to have a higher metallicity.

Quasar NLR abundances have also been studied.  \citet{Nagao_2006_AGN_NLR_Z} compared NLR quasar rest-frame ultraviolet emission to photoionization models, finding a lack of metallicity evolution and a correlation between luminosity and metallicity, as is the case for BLRs.  \citet{Dors_2014_AGN_Z} present a metallicity indicator for AGN NLRs using UV lines, and apply the diagnostic to high-redshift quasars.  \citet{Du_2014_NLR_BLR_Z} studied the relationship between BLR and NLR metallicities, finding that abundances in the different regions are correlated.

The abundances in the Seyfert nuclei of local star-forming galaxies may be estimated using radial abundance gradients.  Two Seyfert NLR abundance calibrations were developed by \citet{Storchi-Bergmann_1998_AGN_Z} by comparing photoionization model predictions to nuclear abundances extrapolated from measured \HII\ region abundances in seven Seyfert galaxies.  Likewise, \citet{Dopita_2014_S7_I} used \HII\ region abundances extrapolated to the nucleus to calculate the gas-phase metallicity of the Seyfert nucleus in the galaxy NGC\,5427.

An overview of broad emission line, broad absorption line, and narrow absorption line abundance diagnostics for QSOs is provided by \citet{Hamann_Ferland_1999_QSO_Z}.  A review of the smaller selection of diagnostics available for NLR abundances is provided by \citet{Dors_2015_AGN_Z}.

Work on measuring NLR abundances is continuing.  A new diagnostic based on the N2O2 index was developed by \cite{Castro_2017_AGNZ}.  The authors vary metallicity and ionization parameter in their photoionization model grids, but do not consider variations in the nebular pressure or the ionizing spectrum.   \citet{Dors_2017_NLR_N_O} measure N and O abundances in nearby Seyfert NLRs, finding the variation of N/O with O/H to be consistent with \HII\ regions.

Despite the recent progress, measurements of NLR abundances remain less routine than is the case for \HII-region measurements.  An example of the ongoing challenges in measuring NLR abundances is the recent work of \cite{Kawasaki_2017_lowZAGN}, who investigated using optical diagnostic diagrams to identify low-metallicity AGN, but did not attempt to measure the metallicity of the objects in their sample.

The historical difficulty in reliably and systematically measuring chemical abundances in active galaxies is a major hindrance to understanding how an AGN may influence the distribution of metals in its host.  In particular, the chemical abundance set, the spectral shape of the ionizing radiation field, the ionization parameter and the gas pressure all affect the optical narrow-line emission signatures of AGN \citep[e.g.][]{Groves_2004_dusty_models}.  In this work we present a grid of photoionization models and analysis code that for the first time is able to measure AGN metallicity while taking into account (and simultaneously measuring) the ionization parameter, pressure, and the hardness of the ionizing continuum.

The shape of the ionizing continuum is a very interesting property in its own right, because this most important part of the SED in the extreme ultraviolet (EUV) is impossible to directly observe, and because knowledge of the ionizing spectrum is essential in order to understand the effects of radiative mode AGN on surrounding gas.  To date it has not been possible to make robust and systematic measurements of the ionizing radiation field.  The radiative hardness may be inferred indirectly using emission-line ratios \citep[e.g.][]{Krolik_Kallman_1988Seyfert_spectrum, Dietrich_2005_EUVspectrum, Kraemer_2008_N4151, Dopita_2014_S7_I}, an approach that we take here in our comparison of emission line fluxes to photoionization models.

The ionization parameter is also an interesting quantity in ENLRs, because the spatial distribution of the ionization parameter gives insights into the effects of a radiative mode AGN on the density structure of the ISM.  The interplay between radiation pressure and gas pressure regulates the structure of individual NLR clouds \citep[e.g.][]{Dopita_2002_Prad} and the NLR on large scales \citep[e.g.][]{Stern_2014_NLR_RPC}.  Recently \citet{Davies_2016_rad_press} studied four Seyfert galaxies with integral field spectroscopy and inferred high, spatially-uniform ionization parameters in the nuclear regions of two objects, arguing that in these regions radiation pressure dominates over gas pressure.

We additionally measure the ISM pressure in ENLRs, although our results are the least sensitive to pressure amongst the four parameters we consider.  The pressure is nevertheless also vital to understanding AGN feedback -- calculations of mass loading and hence energy in outflows are sensitive to the ISM density, which is primarily determined by the pressure \citep[e.g.][]{2011_MullerSnachez_AGN_outflows, Kakkad_2016_SINFONI_outflows}.

The advances we present here are possible through the use of a Bayesian analysis technique, which allows us to efficiently constrain four parameters simultaneously when comparing observations to a grid of photoionization models.  We present a new code, NebulaBayes\footnote{NebulaBayes is publicly available; see the Appendix for more information}, that is heavily inspired by the code IZI \citep{Blanc_2015_IZI} and generalizes its capabilities.  The new code is agnostic to the parameters of the input photoionization model grid, and works with an arbitrary number of dimensions -- limited only by calculation time and available memory for the interpolated $n$-dimensional emission-line flux grids.  We hope that NebulaBayes will help to solve some of the problems discussed above by providing a general and comprehensive means of comparing emission-line fluxes to theory.

We use NebulaBayes to study IFU data from the {\it Siding Spring Southern Seyfert Spectroscopic Snapshot Survey} (S7).  We analyse a sample of Seyfert galaxies with ENLRs covering a substantial portion of the field of view: ESO\,138-G01, IC\,5063, Mrk\,573, and NGC\,2992, as well as a normal star-forming galaxy, NGC\,4691.

In the following section (Section~\ref{sec:Bayes}) we present the background and theory behind the NebulaBayes code.  In Section~\ref{sec:Grids} we describe our NLR and \HII\ region photoionization model grids, and in Section~\ref{sec:Obs_data} we describe the S7 survey, the sample selection and the processing of the observational data.  Section~\ref{subsec:NB_method} contains details of how NebulaBayes was applied to the data.  The results are presented and discussed in Section~\ref{sec:Results}, with some additional discussion in Section~\ref{sec:more_discussion}.  Our conclusions are listed in Section~\ref{sec:Conclusions}.

\section{Bayesian analysis method} \label{sec:Bayes}

\subsection{Background} \label{subsec:intro_Bayes}
Physical parameters such as metallicity have traditionally been measured from emission line data using specific diagnostics.  Example metallicity diagnostics include $R23$ \citep{Pagel_1979_R23, Pilyugin_Thuan_2005_R23}, N2O2 \citep{Kewley_Dopita_2002_Z} and N2 \citep{Denicolo_2002_N2, Pettini_Pagel_2004} among many others; a summary of the calibration methods and diagnostics is provided in the introduction of \citet{Blanc_2015_IZI}.  The diagnostics are each individually calibrated and use a small number of observed emission line fluxes, so cannot take advantage of additional line fluxes when they are available.  The individual diagnostics have historically been heavily favored over techniques that allow for general comparisons between observations and theory.

A means of improving this situation using Bayesian inference was presented by \citet{Blanc_2015_IZI} with the IDL code IZI.  This code simultaneously constrains the ionization parameter and metallicity when provided with any set of observed \HII\ region (or star-forming galaxy) emission-line fluxes and corresponding uncertainties.  The calculations in IZI use 2D arrays (with dimensions corresponding to metallicity and ionization parameter) to examine the relative probability of different models over a large parameter space.

The Bayesian inference method compares observations with theory {\it directly}, so the parameter constraints are conceptually similar to the results of simultaneously calibrating and applying bespoke diagnostics.  A set of observed emission line fluxes is compared to a general grid of photoionization models using a technique that is agnostic to the parameters of interest and constrains the parameters in a systematic, general and mathematically clear manner.  The maximum possible amount of information may be included -- all observed emission line fluxes with their errors and any quantified pre-existing knowledge (i.e.\  priors) are propagated into the final parameter estimates and uncertainties.

A joint probability density function (PDF) resulting from a Bayesian analysis can be visually and quantitatively interrogated across the parameter space to determine constraints, uncertainties and correlations.  Underlying degeneracies and pathological behavior may be revealed by inspecting likelihood PDFs.  Priors may then be applied to select and further constrain solutions using knowledge of the physical system.

A Bayesian analysis requires calculating the posterior PDF (described in the following section).  Algorithmic sampling methods such as those involving Markov Chain Monte Carlo (MCMC) are commonly used to build up a picture of the posterior PDF using a step-by-step probabilistic exploration of the parameter space.  Unfortunately a comparison of emission-line fluxes to predictions using a sampling approach would require running a photoionization model for each visited point in the parameter space.  The run-time of a single typical photoionization model (on the order of one to a few minutes) is too long for this approach to be desirable, considering that the data from the model runs could not be reused in general, and that modern instruments may produce many hundreds of spectra in a single observation.

However, IZI uses an approach that is able to compare observed emission-line fluxes to predictions using Bayesian inference without requiring a full run of a sampling algorithm for each spectrum.  The method uses a pre-calculated (hence re-usable) grid of photoionization model predictions.  The relative probability of a set of parameters describing the observed nebula is directly calculated by comparing the observed fluxes to predicted fluxes found by interpolating between the model grid points.  The probability is calculated across the entire parameter space, to produce an $n$-dimensional joint PDF.  This is a `brute force' approach to Bayesian inference that involves manipulation of arrays representing the entire parameter space, so is only applicable when the parameter space has low dimensionality.

Another recent work applying a similar approach is that of \citet{ValeAsari_2016_BOND}.  With their code {\sc BOND} the authors use three-dimensional grids with parameters for the abundance ratios O/H and N/O, as well as the ionization parameter.  A related strategy is used by \citet{Perez-Montero_2014_HII-CHI-MISTRY} in the code {\sc HII-CHI-MISTRY}, which determines parameter values by weighting models in a grid by the reciprocal of the $\chi^2$.  \citet{Bianco_2016_PyMCZ} take a less similar approach in their code PyMCZ, measuring nebular oxygen abundances using specific calibrators and using Monte Carlo sampling to estimate the errors.

Our new code NebulaBayes generalizes the methods used in IZI and BOND, providing parameter estimates and statistical errors from observed line fluxes.  A first analysis of ENLR spectra with NebulaBayes was performed by \citet{Merluzzi_2018_light_echo}; the inferred parameters were consistent with the results produced by a more detailed analysis with tailored photoionization models.  The NebulaBayes software is described in detail in the following section and in Appendix~\ref{sec:NB}.

\subsection{Theory} \label{subsec:Theory}
We use Bayesian parameter estimation to determine the probability of model parameter values given observed data.  In this section we follow Section~2 of \citet{Blanc_2015_IZI}.

At the heart of the method is Bayes' theorem, which is readily derived using basic probability theory and reads as follows:
\begin{equation}
p({\boldsymbol \theta} | D, M') = \frac{ p({\boldsymbol \theta} | M') \, p(D | M',{\boldsymbol \theta})}{p(D | M')} .
\label{eq:Bayes_theorem}
\end{equation}
The quantity $p({\boldsymbol \theta} | D, M')$ is the `posterior' and is the probability of a particular set of model parameter values ${\boldsymbol \theta}$ being `true' given the available observed data $D$ and a model $M'$.  We want to evaluate the posterior for many sets of values ${\boldsymbol \theta}$ in order to draw conclusions about the physical system that gave rise to the observed data.  The quantity $p(D | M',{\boldsymbol \theta})$ is the hypothetical `likelihood' of the observed data given the model $M'$ and the values of the parameters ${\boldsymbol \theta}$.  The `prior' $p({\boldsymbol \theta} | M')$ is assigned carefully by hand and describes the probability of the parameter values considering the model and all previously known information.  The normalization $p(D | M')$ is found by normalizing the posterior and is unimportant in parameter estimation problems.

We assume that emission line measurements follow a Gaussian distribution centered at the `true' value.  Let $f_i$ be an emission-line flux measurement (for line $i$), $e_i$ the associated measurement error, $M'$ a particular model with a set of model parameter values ${\boldsymbol \theta}$, and $f'_i({\boldsymbol \theta})$ the predicted emission-line flux associated with the values ${\boldsymbol \theta}$.  Then the probability of measuring a particular value of $f_i$ given the assumed model $M'$ and ${\boldsymbol \theta}$ is
\begin{equation}
p(f_i | M', {\boldsymbol \theta} ) \propto \frac{1}{e_i} \exp \left[ - \frac{(f_i - f'_i({\boldsymbol \theta}) )^2}{2e_i^2} \right] .
\label{eq:naive_prob}
\end{equation}

Equation~\ref{eq:naive_prob} assumes that $f'_i({\boldsymbol \theta})$ is a perfect model prediction based on the input parameters.  In reality the calculation of predictions from a set of parameters ${\boldsymbol \theta}$ uses physical simulations and hence is associated with inherent uncertainty.  Following \cite{Blanc_2015_IZI}, we make an attempt to account for the modelling uncertainty by including systematic errors with the predictions.  The measurement errors and systematic model errors are independent, so we simply sum the variances to give a total variance $V_i = e_i^2 + \epsilon^2 \, f'_i({\boldsymbol \theta})^2$ where $\epsilon$ is the estimated uniform fractional systematic error in the model predictions.  This produces an appropriately broader Gaussian distribution:
\begin{equation}
p(f_i | M', {\boldsymbol \theta} ) \propto \frac{1}{\sqrt{V_i}} \exp \left[ - \frac{(f_i - f'_i({\boldsymbol \theta}) )^2}{2 V_i} \right] ,
\label{eq:usual_likelihhod_contribution}
\end{equation}

In the case of an upper bound on the observed line flux, we assume that only $e_i$ is supplied, because $f_i < e_i$ and the $f_i$ measurement was discarded.  In this case we are interested in the probability of measuring $f_i < e_i$, assuming that $f'_i({\boldsymbol \theta})$ is the true flux:
\begin{equation}
p(f_i < e_i | M', {\boldsymbol \theta}) = A \frac{1}{\sqrt{V_i}} \int_0^{e_i} \exp \left[ - \frac{(f_i - f'_i({\boldsymbol \theta}) )^2}{2 V_i} \right] \mathrm{d}f_i
\label{eq:UB_unreduced}
\end{equation}
for a constant A.  Substituting $t = (f_i - f'_i({\boldsymbol \theta})) / \sqrt{2 V_i}$ and using the definition of the error function $\mathrm{erf}(x) = 2 \, \pi^{-0.5} \int_0^x e^{-t^2} \mathrm{d}t$, we obtain
\begin{equation}
p(f_i < e_i | M', {\boldsymbol \theta}) = B \left[ \mathrm{erf}\left[\frac{f'_i({\boldsymbol \theta})}{\sqrt{2 V_i}}\right] - \mathrm{erf}\left[\frac{f'_i({\boldsymbol \theta}) - e_i}{\sqrt{2 V_i}}\right] \right]  \label{eq:UB_reduced}
\end{equation}
for a constant B.

The likelihood, which is the PDF describing the probability of obtaining the emission line measurements given the assumed model and set of parameter values, is
\begin{equation}
p(D | M', {\boldsymbol \theta} )  \propto \prod_{i=1}^{m} \; p_i
\label{eq:full_likelihood}
\end{equation}
where there are $m$ measured emission lines, the data $D$ is the full set of measured emission line fluxes and errors $\{f_i, e_i\}$ (with zero or more `upper bound' measurements for which $f_i$ is excluded), and the contribution of a single line $p_i$ is of the form given in either Equation~\ref{eq:usual_likelihhod_contribution} or Equation~\ref{eq:UB_reduced}.

In Equation~\ref{eq:full_likelihood} it is assumed that the $p_i$ contributions are independent.  In practice this is not the case, with correlations expected to arise between the $p_i$ because the line fluxes are all normalized to a reference line (\Hb\ in this work), and because of systematic issues such us uncertainties in reddening corrections.

Much of the constraining power of emission-line observations comes from specific sensitive flux ratios.  Hence a treatment in which we compare all possible ratios to predictions might be expected to constrain the parameter space more effectively than using individual line fluxes as the data points.  However the expressions above would need to be much more complicated in this case, because the error distributions on the ratios would not in general be close to Gaussian, and the likelihood contributions of the ratios would not be close to independent (each flux would appear in multiple ratios), so Equation~\ref{eq:full_likelihood} would need to account for correlations.  Nevertheless we are able to take advantage of the constraining power of specific line ratios by using them in priors (Appendix~\ref{sec:NB}).

The code NebulaBayes calculates the likelihood PDF of Equation~\ref{eq:full_likelihood} over the whole parameter space covered by the photoionization models.  The likelihood is then multiplied by a prior chosen by the user (also calculated over the entire parameter space), to produce the full $n$-dimensional joint posterior PDF as per Equation~\ref{eq:Bayes_theorem}.

We note that emission lines are weighted equally in Equation~\ref{eq:full_likelihood}.  Given a photoionization model grid, a user of NebulaBayes may make only a small number of choices which are able to modify the results:
\begin{enumerate}
	\item The choice of the prior
	\item The choice of the set of observed emission-lines to use
	\item The systematic error on the predicted fluxes, $\epsilon$
    \item Choices related to how to deredden the observed data
\end{enumerate}
This small number of options is sufficient to produce significantly differing results -- in particular, judicious selection of the prior and of the set of lines to use is essential.

The implementation of NebulaBayes and the straightforward methods of obtaining the code are described in Appendix~\ref{sec:NB}.

\vspace{2cm}
\section{Photoionization models} \label{sec:Grids}

NebulaBayes includes predicted emission-line fluxes over two photoionization model grids, which are also the grids used in this work.  The two grids are for \HII\ regions and for NLRs.  All models were produced by the MAPPINGS photoionization code, described in Section~\ref{subsec:MAPPINGS}.

For setting abundances in the models we use the work of \citet{Nicholls_2017_zeta}, which involves a more sophisticated scaling of abundances with metallicity than the uniform scaling that is commonly applied.  This scaling accounts for the changing contributions of primary and secondary nitrogen with metallicity and for the changing ratio of $\alpha$-process elements to iron-peak elements as a function of metallicity.

\subsection{MAPPINGS} \label{subsec:MAPPINGS}
The MAPPINGS photoionization and radiative shock wave modeling code has been developed over three decades.  Already in its first iteration \citep{Binette_1985_MAPPINGS} it was capable of modeling plasmas which are in neither photoionization nor collisional equilibrium (e.g.\ shocks).  The code was subsequently improved with the addition of new physical processes, ions, and atoms \citep{Sutherland_Dopita_1993_cooling_func}, dust heating \citep{Dopita_Sutherland_2000_Dust}, non-equilibrium dust heating and infrared emission \citep{Groves_2004_dusty_models,Groves_2006_IR,Dopita_2005_starburst}, and treatment of non-equilibrium ($\kappa$-distribution) electron energies \citep{Dopita_2013_kappa, Nicholls_2012_kappa}, among many other improvements.  The latest version of the code \citep[MAPPINGS\,V;][]{Sutherland_Dopita_2017_shocks} takes advantage of new, highly detailed atomic data, and tracks more than $8 \times 10^4$ cooling and recombination emission lines up to densities of order $10^{12}$\,cm$^{-3}$.  The photoionization model grids described below were computed with MAPPINGS version 5.1.

\subsection{Narrow line region model grid} \label{sec:NLR_grid}
The NLR grid uses an {\sc oxaf} ionizing spectrum \citep{Thomas_2016_oxaf}, which has three parameters: the energy of the peak of the accretion disk emission $E_{\rm peak}$, the photon index of the inverse Compton scattered power-law tail $\Gamma$, and the proportion of the total flux that goes into the non-thermal tail, $p_{\rm NT}$.  The two parameters $\Gamma$ and $p_{\rm NT}$ are somewhat anti-correlated, because decreasing the hardness of the power-law tail by increasing $\Gamma$ has a similar effect (at the most relevant EUV energies and at fixed $p_\mathrm{NT}$) as scaling up the power-law tail with $p_\mathrm{NT}$.  The {\sc oxaf} model does not account for the soft X-ray excess emission; \cite{Thomas_2016_oxaf} showed that including the soft excess does not have a major effect on predictions of the strong optical emission lines.  In addition, a soft X-ray excess may arise from the Compton-heated photoionized gas close to the nucleus, rather than being an intrinsic part of the EUV spectrum of the nucleus itself.

For our NLR grid, the only {\sc oxaf} parameter we vary is $E_{\rm peak}$, because it is a computational necessity to reduce the number of dimensions.  The other parameters are fixed to the fiducial values $\Gamma = 2.0$ and $p_\mathrm{NT} = 0.15$.  Our final NLR grid has four parameters or dimensions: metallicity, ionization parameter, pressure, and $E_{\rm peak}$.

The grid was run using the following configuration:
\begin{itemize}
	\item Oxygen abundances sampling the 12 values  $\log {\rm [O/H]} =$~-1.70, -1.10, -0.70, -0.40, -0.15, 0.00, 0.11, 0.23, 0.32, 0.40, 0.48, and 0.54, using the oxygen-based standard abundance scaling from \citet{Nicholls_2017_zeta}, with the `local galactic concordance' reference abundance values.  In this scale the Solar oxygen abundance is $12 + \log {\rm O/H}$ = 8.76 and our abundances sample 0.02 to 3.5~Solar.
	\item Ionization parameter at the inner edge of the modeled nebula sampling 11 values in the range $-4.2 \leq \log U \leq -0.2$, uniformly spaced at a logarithmic interval of 0.4\,dex
	\item Initial gas pressures sampling 12 values in the range $4.2 \leq \log P/k$\;(cm$^{-3}$\;K)\;$\leq 8.6$ uniformly at a logarithmic interval of 0.4\,dex.  The total pressure is given by the sum of the gas pressure and the radiation pressure, such that the total pressure increases through the model as the radiation is absorbed.
	\item Values of $E_{\rm peak}$ sampling 6 values in the range $-2.0 \leq \log$\,\Ep$\;\mathrm{(keV)} \leq -0.75$ uniformly at a logarithmic interval of 0.25\,dex
	\item A plane parallel geometry
	\item A set of depletions onto dust grains based on iron being 97.8\% depleted ($\log ({\rm Fe_{\,free} / Fe_{\,total}}) = -1.5$; \citet{Jenkins_2009_depletions, Jenkins_2014_depletions})
	\item No dust destruction
	\item An equilibrium, Maxwell-Boltzmann distribution of electron energies ($\kappa = \infty$) \citep{Nicholls_2012_kappa}
\end{itemize}
The total number of gridpoints in the NLR grid is 9504.  We note that the predicted intrinsic Balmer decrements ($F_{\mathrm{H}\alpha} / F_{\mathrm{H}\beta}$) range between 2.79 and 3.54, with a median and 75th percentile of 2.94 and 3.06 respectively.  These values compare to the commonly assumed ratio of 2.86 for Case B recombination.

Predicted fluxes for 119 emission lines are included for each point in the grid.  The list of lines was selected from a low-abundance \HII\ region, a high-abundance \HII\ region, and a high-excitation NLR MAPPINGS model.  The selected lines all have a flux above 1\% of \Hb\ in at least one of the models, have a wavelength between the Lyman limit and 20\,$\mu$m, and arise from a species with an ionization potential below 100\,eV.  Lines from iron and magnesium were excluded because we do not allow dust destruction in the MAPPINGS models, so the relevant fluxes are very inaccurate.  We also exclude higher-order recombination lines (e.g.\ H8, Pa7, etc.) that are difficult to predict correctly without a very complete recombination-cascade solution as a function of density.  Summed fluxes are included for some commonly considered doublets for convenience.  The line list is distributed with NebulaBayes.

\subsection{\HII\ region model grid} \label{sec:HII_grid}

The \HII\ region grid has three parameters or dimensions: abundance, ionization parameter, and pressure.  The grid was run with the same abundances, pressures, depletions and MAPPINGS settings as the NLR grid described above.  The ionization parameter was varied over $-4.0 \leq \log U \leq -2.0$ in increments of 0.25\,dex.  The only other difference was the form of the ionizing spectrum, which varied with metallicity (but not freely with its own parameter).  This was to ensure that the stellar and nebular metallicities are approximately matched.  Perfect matching of stellar and nebular metallicities is not possible because the stellar atmosphere modelers use older Solar reference abundances, as well as a restricted set of initial abundances.

The ionizing spectra were derived using the {\sc slug2} \citep{Krumholz_2015_SLUG2} stellar population synthesis code with the following settings:
\begin{itemize}
	\item The `galaxy' (continuous star formation) mode
	\item A single snapshot of the spectrum after $10^7$ years
	\item A star formation rate of 0.001\,$M_\odot \,$yr$^{-1}$.
	\item A Chabrier IMF
	\item Default Starburst\,99 \citep{Leitherer_1999_SB99} spectral synthesis mode
	\item The Padova stellar tracks with thermally pulsing AGB stars
	\item The five explicitly calculated available metallicities, which were $Z = 0.0004, 0.004, 0.008, 0.02, 0.05$, where $Z = 0.02$ is Solar
\end{itemize}
The spectra for the five different $Z$ values were interpolated to the 12 oxygen abundances used in the NLR grid.  The oxygen abundance-metallicity scaling of \citet{Nicholls_2017_zeta} was used, and the spectra were linearly interpolated in $F_\lambda$ space.  The exception was that the spectra for 3.0 and 3.5\,$Z_\odot$ were the same as for 2.5\,$Z_\odot$, to avoid extrapolation.  The grid was run with the 12 nebular abundances, with the appropriate spectrum chosen depending on metallicity.

The total number of gridpoints in the \HII\ region grid is 1296.  The list of included emission lines was the same as for the NLR grid except for 26 excluded higher-ionization lines that had no fluxes above $10^{-5}$ of \Hb.  The predicted intrinsic Balmer decrements ($F_{\mathrm{H}\alpha} / F_{\mathrm{H}\beta}$) range between 2.81 and 3.59, with a median and 75th percentile of 2.93 and 3.01 respectively; the distribution is very similar to that for the NLR grid.

A fully self-consistent modeling of \HII\ region emission line spectra is a major challenge at present, because the predictions are very sensitive to the assumptions that are used in generating the ionizing spectra.  The most important regions of the ionizing spectrum are those defined by the ionization potentials of the key nebular species, and the ionizing flux in these regions varies significantly with the chosen stellar tracks/atmospheres, model assumptions (e.g. stellar rotation, dredge up of helium, mass-loss rates, effects of binarity) and the interpolation schemes used.  It is necessary to match the stellar abundances to the nebular abundances for physical consistency, but inconsistent abundances are used between stellar track and atmosphere models.  Also, at present the parameter space coverage of stellar evolutionary tracks and model stellar atmospheres is inadequate in both extent and sampling frequency.  In this work we make a reasonable attempt to approximately match stellar and nebular abundances; fully self-consistent \HII-region models are currently under development (PI Sutherland).

In a future version of MAPPINGS we intend to incorporate the diffuse super-soft thermal X-ray continuum arising from thermalization of the fast stellar wind through shocks.  This may improve the predictions of some lines that arise in the partially-ionized zone.

\section{Observational data} \label{sec:Obs_data}

\subsection{S7} \label{sec:S7}
The \emph{Siding Spring Southern Seyfert Spectroscopic Snapshot Survey} \citep[S7;][]{Dopita_2015_S7_II} is an optical integral field spectroscopy survey using the Wide Field Spectrograph (WiFeS) mounted on the ANU 2.3~m telescope at Siding Spring Observatory, Australia.  The S7 was carried out from 2013 -- 2016 and contains a sample of 131 local ($z < 0.02$) galaxies, including approximately 77 Seyfert~2 galaxies, 18 Seyfert~1 galaxies, 27 LINERs, and 11 star-forming-only galaxies.  The WiFeS field of view of 38~$\times$~25~arcsec$^2$ covers the centre of each galaxy.  The resulting data cubes consist of a grid of 1~$\times$~1~arcsec$^2$ spatial pixels (`spaxels'), an angular resolution comparable to the median seeing across the survey of 1.5~arcsec.  The design and performance of WiFeS are discussed in \citet{Dopita_2007_WiFeS_I} and \citet{Dopita_2010_WiFeS_II}.

The spectral resolution of the S7 data is $R = 7000$ in the red (FWHM $\sim 40$\,km\,s$^{-1}$ over $540 - 700$\,nm) and $R = 3000$ in the blue (FWHM $\sim 100$\,km\,s$^{-1}$ over $350 - 570$\,nm), which is higher than comparable surveys and permits analysis of separate velocity components in the emission lines.  The data have a high spatial resolution at the very low redshift of the sample, which also allows separation of extended regions ionized by the AGN, star formation, and shocks.

Ongoing analysis of the S7 data motivated the development of the NebulaBayes code.  We required a general method to constrain photoionization model parameters not just in \ion{H}{2} regions, but also in extended AGN narrow-line regions, in high-ionization nuclear coronal line clouds in AGN, and potentially also in shock-excited regions and LINER nuclei.  The approach is required to be computationally efficient when applied to hundreds or thousands of spectra.

\subsection{Sample selection} \label{sec:sample_selection}
The sample selection was driven by our desire to minimize contamination from non-Seyfert excitation and to maximize the area in the S7 field of view which could be analyzed.  These objectives ensured that we could usefully compare our data to a NLR model grid and make strong inferences regarding the spatial distributions of physical parameters.  

\begin{figure*}
	\centering
	\includegraphics[width=1.0\textwidth]{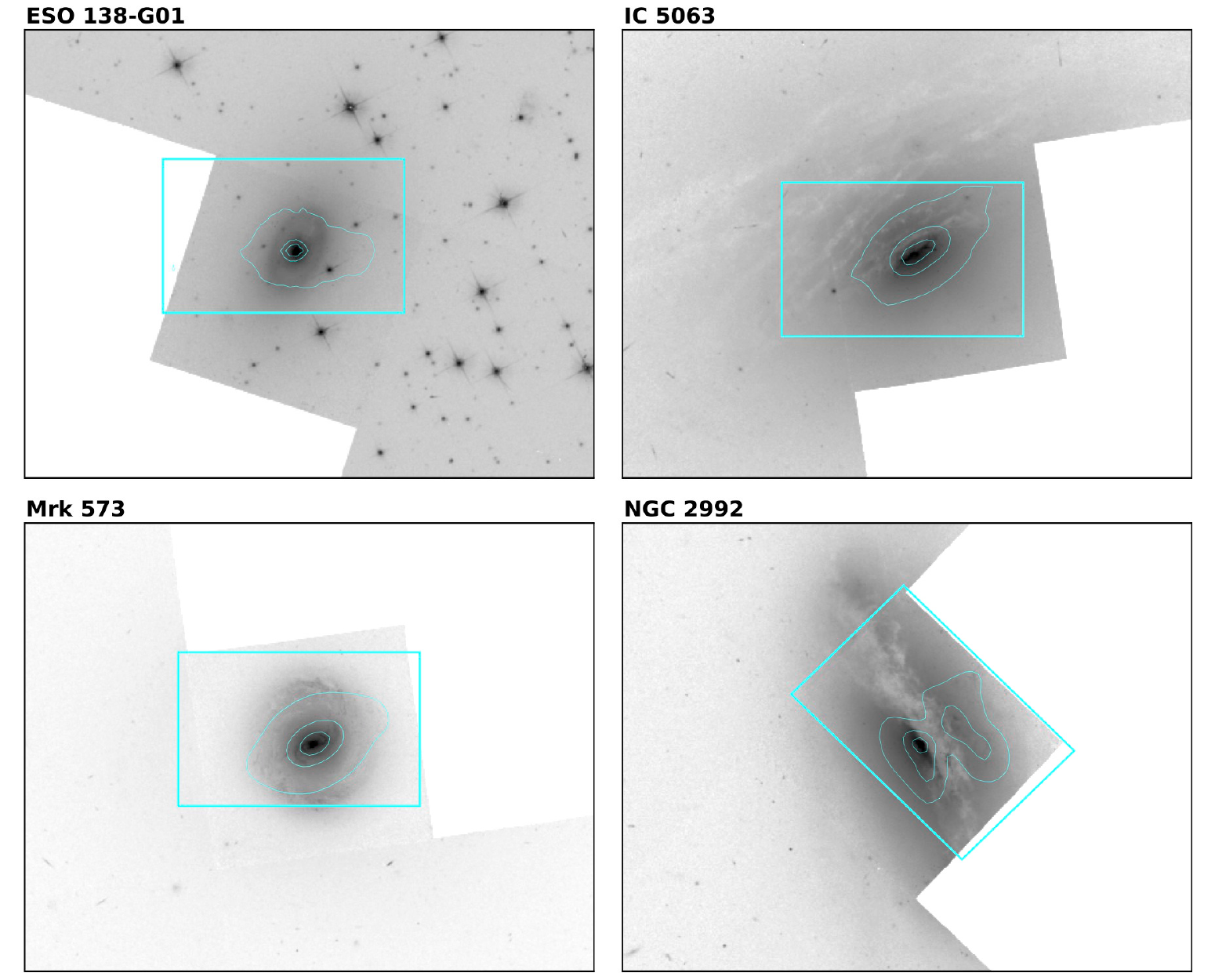}
	\caption{Archival HST WFPC2 F606W wide-band optical imaging \citep{Malkan_1998_HST_AGN} for the four nearby Seyfert galaxies studied in this work.  The cyan rectangles approximately show the $25 \times 38$~arcsec$^2$ field of view of WiFeS.  Logarithmically-spaced contours reflect the \OIII\ flux map measured from single-component fits to the S7 IFU data.  North is toward the top of the page and east is to the left.\\ \label{fig:HST}}
\end{figure*}

\begin{figure*}
	\centering
	\includegraphics[width=0.87\textwidth]{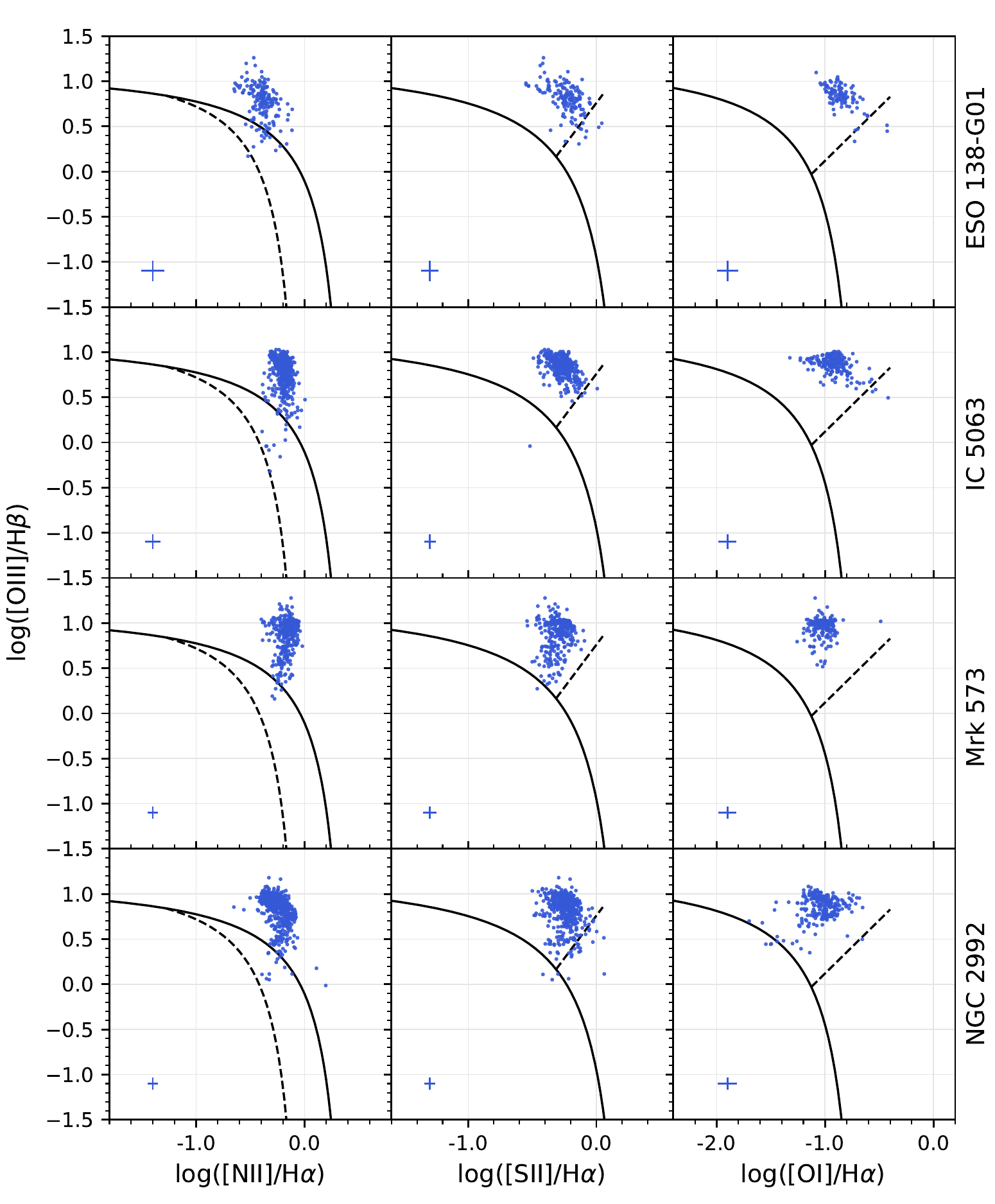}
	\caption{Optical diagnostic diagrams \citep[BPT/VO diagrams;][]{BPT_1981,1987VO} showing spatially-resolved emission-line flux ratios for the four selected Seyfert galaxies.  The solid line in each panel is a theoretical separation between nebulae associated with star formation (below) and higher-excitation nebulae \citep[above;][]{Kewley_2001_starburst}.  The dotted line in the leftmost panels is an empirical variant of the same division \citep{Kauffmann_2003_AGN}.  The dotted line in each panel of the rightmost two columns separates LINERs (below) and Seyferts \citep[above;][]{Kewley_2006_AGN_hosts}.  Each datum is derived from single-component Gaussian line fits for a 1~arcsec square spatial pixel.  All line fluxes have S/N~$>$~2, and median error bars are shown in the bottom-left of each panel.  Seyfert excitation appears to generally dominate over other sources of excitation in the observed regions of these four galaxies. \label{fig:BPTVO}}
\end{figure*}

Four Seyfert galaxies were selected from the S7 sample: ESO\,138-G01, IC\,5063, Mrk\,573, and NGC\,2992.  Images of these objects are shown in Figure~\ref{fig:HST}, which illustrates the S7 field of view.  These galaxies are the examples that most clearly meet the following criteria:
\begin{enumerate}
	\item They are Seyfert galaxies with ENLRs extending across the majority of the length and width of the WiFeS field of view 
	\item In 1-component line fitting to the unbinned data, the excitation across the WiFeS field of view was dominated by emission-line ratios falling in the `Seyfert' regions of the [\ion{N}{2}] and [\ion{S}{2}] optical diagnostic diagrams (Figure~\ref{fig:BPTVO}), and in particular, there was no significant evidence for excitation by O stars.
	\item The signal-to-noise (S/N) was generally sufficient for a high-quality NebulaBayes analysis, i.e.\ spatially binning to a target S/N of 100 or 150 in \Ha\ flux resulted in a satisfactory number of spatial elements being retained
\end{enumerate}

A handful of other galaxies in the S7 sample had emission dominated by Seyfert excitation, but over a smaller area of the field of view compared to the selected galaxies (e.g.\ ESO\,103-G35, IRAS\,01475-0740).  Approximately 30\% of the S7 galaxies featured regions of multiple spaxels with Seyfert-dominated emission, but were excluded from our sample because non-Seyfert regions were also present (usually with `composite' and/or \HII-region classifications).

\begin{figure*}
	\centering
	\includegraphics[width=1.0\textwidth]{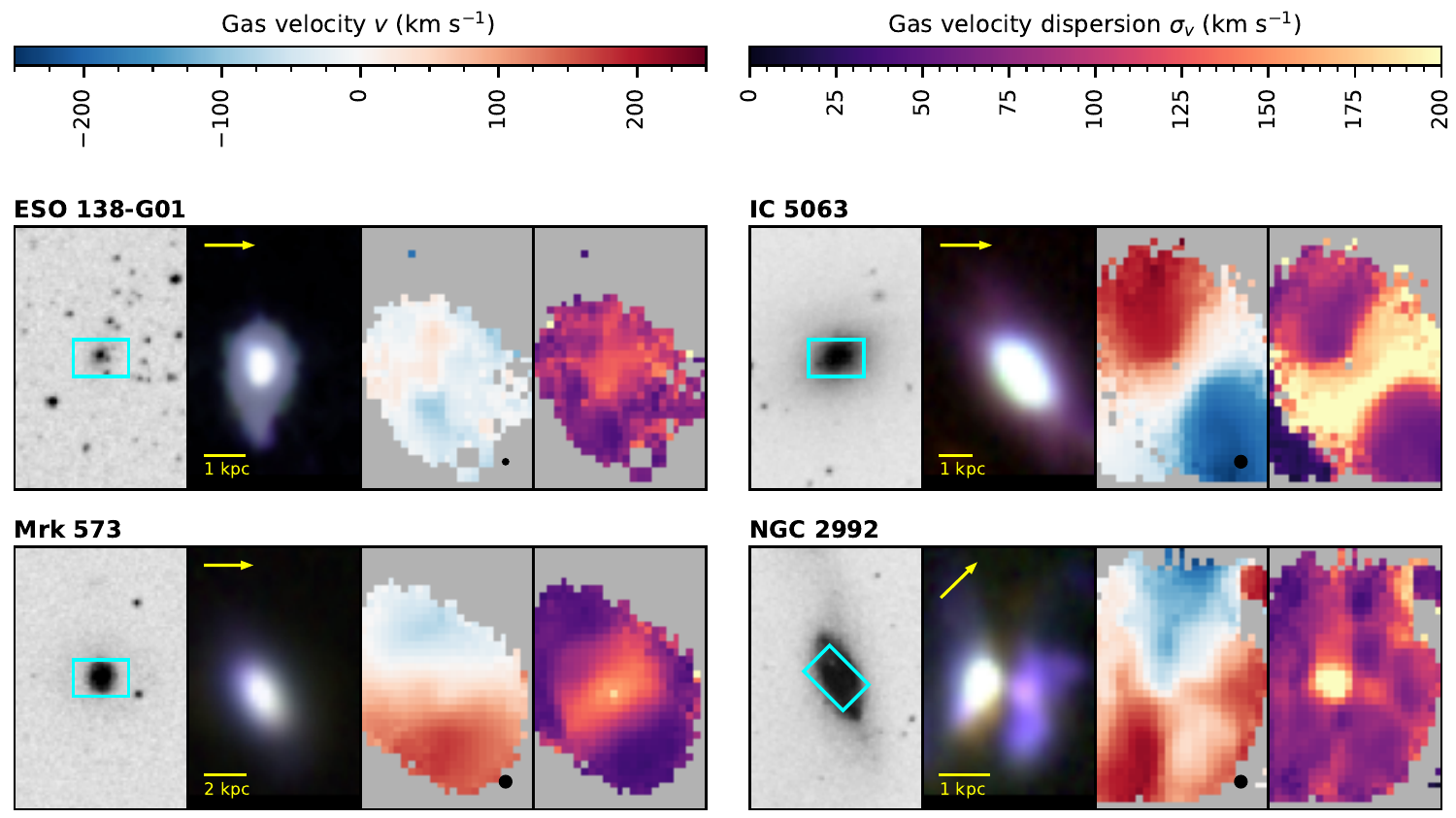}
	\caption{Maps for the four selected `pure Seyfert' S7 galaxies.  The panels from left to right: a) A DSS2 r-band image; north is upwards and the cyan rectangle shows the $25 \times 38$\,arcsec$^2$ WiFeS field of view.  b) A composite image illustrating the ionization state of the gas.  The red channel is \Ha, the green channel is \lNII, and the blue channel is \lOIII.  ENLRs are generally purple (strong \lOIII\ and generally high \NIIHa), whereas \HII\ regions would appear red or gold.  The yellow arrow indicates the direction of north in the last three panels.  c) The ionized gas line-of-sight velocity from single-component emission-line fits.  A black circle in the lower-right shows the seeing FWHM.  d) A corresponding map of line-of-sight velocity dispersion.  The velocity dispersion maps show evidence of outflows in the ENLRs, especially along the minor axis of IC\,5063.\label{fig:Basic_maps}}
\end{figure*}

The selection based on BPT/VO diagrams cannot guarantee that the observed emission is only from Seyfert photoionization.  There are two key reasons that we expect contamination.  Firstly, the classifications of Figure~\ref{fig:BPTVO} are approximate: for example, the LINER-Seyfert distinction is inexact \citep{Kewley_2006_AGN_hosts} and grids of fast shocks overlap the Seyfert region of the diagnostic diagrams \citep{Allen_2008_shocks, Dopita_Sutherland_1996_fast_shocks}.  Secondly, the nature of Seyfert galaxies and the IFU data strongly suggests that some contamination will be present.  Ubiquitous outflows in Seyfert galaxies are associated with shocks, and the gas that fuels the black hole could equally be forming ionizing young stars.  We do not resolve individual nebulae, so our spectra are luminosity-weighted sums of both Seyfert-photoionized and contaminating emission over varying projected areas (spatial scales are shown in Figure~\ref{fig:Basic_maps}).
	
There is some evidence of contamination in Figure~\ref{fig:BPTVO}.  Contaminating spectra will make varying contributions to different line fluxes and will generally skew diagnostic line ratios.  Figure~\ref{fig:BPTVO} shows some apparent `mixing' down into the `composite' region of the optical diagnostic diagrams for all four galaxies.  There are also hints of mixing with `LINER-like' spectra (due to shocks or photoionization with a low ionization parameter) on the \SII\ diagram for all galaxies except Mrk\,573.  We discuss how spectral contamination may have affected our results in Section~\ref{sec:Results} and Section~\ref{sec:more_discussion}.

The sample properties are given in Table~\ref{table:sample}.  As well as the four `pure Seyfert' galaxies, we include the `pure-\ion{H}{2}' galaxy NGC\,4691 to demonstrate the application of NebulaBayes to the spectra of star-forming galaxies.  This galaxy was only included in the S7 sample because it was misclassified as a Seyfert in the original \citet{Veron_2010_13ed} catalog.

The HST imaging in Figure~\ref{fig:HST} shows that the morphologies of the four Seyfert galaxies are much more complicated than the HyperLeda morphological classifications in Table~\ref{table:sample} suggest.  ESO\,138-G01 has a stellar ring and Mrk\,573 has spiral arms, so these appear to be disky galaxies.  IC\,5063 and NGC\,2992 feature prominent dust lanes.

The Seyfert galaxies have similar stellar masses, typical for the galaxies included in the full S7 sample (the distribution peaks at $\log M_* / M_\odot = 10.7$).

ESO\,138-G01, IC\,5063 and Mrk\,573 were all observed in photometric conditions, but NGC\,2992 was observed through some cirrus.

\begin{DIFnomarkup}
\begin{deluxetable*}{lDDDcDc}
	\centering
	\tabletypesize{\scriptsize}
	\tablewidth{12cm}
	\tablecaption{The galaxies selected from the S7 sample.   The first four are `pure Seyfert' galaxies that show AGN-dominated optical emission-line excitation across the WiFeS field of view. \label{table:sample}}
	\tablehead{
		\multicolumn{1}{l}{Name} &
		\multicolumn{2}{c}{RA} &
		\multicolumn{2}{c}{Dec.} &
		\multicolumn{2}{c}{Redshift $z$} &
		\multicolumn{1}{c}{Morphology\tablenotemark{1}} &
		\multicolumn{2}{c}{$\log M_* / M_\odot$\tablenotemark{2}} &
		\multicolumn{1}{c}{Excitation type\tablenotemark{3}}
	}
	\startdata
	\decimals
	ESO\,138-G01 & 252.83542 & -59.2364 & 0.0091 & E-S0 & 10.8 & Seyfert 2  \\
	IC\,5063     & 313.00921 & -57.0686 & 0.0113 & S0-a & 10.6 & Seyfert 2  \\
	Mrk\,573     &  25.99079 & +2.3497  & 0.0172 & S0-a & 10.9 & Seyfert 2  \\
	NGC\,2992    & 146.42500 & -14.3264 & 0.0077 & Sa   & 10.5 & Seyfert 2  \\
	NGC\,4691    & 192.05421 & -3.3331  & 0.0037 & S0-a & 9.6  & \ion{H}{2} \\[0.2cm]
	\enddata
	\tablenotetext{1}{From HyperLeda}
	\tablenotetext{2}{Stellar mass estimated as described in \citet{Thomas_2017_S7DR2}, except for NGC\,4691 for which we used Equation\,8 of \citet{Taylor_2011_Mstar} with SDSS $g$ and $i$ magnitudes}
	\tablenotetext{3}{S7 nuclear classification}
\end{deluxetable*}
\end{DIFnomarkup}

In Figure~\ref{fig:Basic_maps} we present maps showing the gas excitation and the kinematics resulting from single-component Gaussian fits to the data.  The figure illustrates physical scales on the galaxies and the seeing FWHM for each observation.

In the following paragraphs we briefly describe the properties of the active galaxies and the features evident in the S7 data.

ESO\,138-G01:  This galaxy hosts a luminous Seyfert~2 nucleus \citep{Fairall_1988_z_survey} and has been previously classified as E/S0 \citep{ESO_atlas}, but appears more disky than elliptical in Figure~\ref{fig:HST}.  The large, asymmetric ENLR has a prominent extension to the west, which is also visible in moderate blueshifts in the velocity map in Figure~\ref{fig:Basic_maps}.  An isolated cloud of \OIII\ emission (too faint to be visible in Figure~\ref{fig:Basic_maps}) is located ${\sim}15\arcsec$ to the east of the nucleus at the edge of the S7 field of view.  The nuclear spectrum boasts a spectacular array of forbidden high-ionization (coronal) emission lines such as [\ion{Fe}{7}]$\,\lambda 6087$ and [\ion{Fe}{10}]$\,\lambda 6375$ \citep{Alloin_1992_ESO138G01}.  Very high ionization lines are also observed in X-ray spectra by \citet{DeCicco_2015_ESO138G01}.  We note that \citet{DeCicco_2015_ESO138G01} measured a soft X-ray photon power law index of $\Gamma = 2.03 \pm 0.14$, an excellent match to the $\Gamma = 2$ we assumed in the photoionization models (Section~\ref{sec:NLR_grid}).

IC\,5063:  A well-studied galaxy exhibiting both Seyfert~2 excitation and a radio jet that powers a spectacular galactic wind \citep[e.g.][]{Sharp_BlandHawthorn_2010_winds, Tadhunter_2014_Nature_IC5063, Morganti_2015_IC5063, Dasyra_2015_IC_5063_SINFONI}.  The biconical, outflowing ENLR has a large opening angle of ${\sim}90$\arcdeg\ and extends at least 30\arcsec\ across the WiFeS field of view.  The nuclear line profiles are extremely complex with multiple independent components.  There is a hidden broad-line region revealed by spectropolarimetry \citep{Lumsden_2004_hiddenBLR}.

Mrk\,573:  A Seyfert~2 galaxy that features a black hole accreting near the Eddington limit \citep{Bian_2007_Edd_ratios, Kraemer_2009_Mark573}, and an ENLR over 8\,kpc in extent that covers most of the WiFeS field of view diagonally from southeast to northwest.  The nucleus shows coronal line emission \citep{Dopita_2015_S7_II}.  HST photometry and spectroscopy led \cite{Fischer_2010_Mrk_573} to argue that an inclined ionization cone intersects the inner spiral arms.  \cite{Rose_2015_AGN_inclination} estimate the nuclear inclination of Mrk\,573 to be $60 \pm 5$ degrees using WISE infrared colors and arguing that the presence of coronal lines is indicative of an intermediate inclination of the obscuring torus.  The excitation of the coronal lines was modeled by \citet{Kraemer_2009_Mark573}.

NGC\,2992:  A disky galaxy interacting with its neighbor NGC\,2993 and showing tidal distortions \citep{Toomre_and_Toomre_1972_tidal}.  Opposing ENLR cones extend with large opening angles either side of the disk, both easily visible due to the high inclination of NGC\,2992 \citep[e.g.][]{Allen_1999_NGC_2992}.  The nuclear spectrum is highly variable and has been observed with and without broad Balmer wings at various times over the last four decades \citep{Ward_1980_NGC_2992, Allen_1999_NGC_2992, Gilli_2000_NGC_2992, Lumsden_2004_hiddenBLR, Trippe_2008_NGC_2992}.  The second panel in Figure~\ref{fig:Basic_maps} is brown along the dust lane; the \OIII\ flux is relatively weak here due to extinction and not because \HII\ regions are dominating the spectrum.

\subsection{Data processing} \label{sec:data_processing}

We use the reduced data cubes from S7 Data Release~2 \citep{Thomas_2017_S7DR2}.  The first post-processing step was to perform a subpixel shift of the blue data cube in order to align it with the red cube -- the PyWiFeS pipeline \citep{Childress_2014_PyWiFeS} differential atmospheric refraction (DAR) correction was inadequate and the final analyses were affected unless this correction was made.  The sizes of the subpixel shifts were determined by eye and the shifts were performed for ESO\,138-G01, IC\,5063 and Mrk\,573.  The corrections were most important for IC\,5063 and Mrk\,573.

The cubes were spatially binned using a Voronoi tessellation \citep{Cappellari_Copin_2003_Voronoi} based on the S/N of the \Ha\ line fluxes in single-component fits to every spaxel.  The target S/N was 100 for ESO\,138-G01 and 150 for IC\,5063, Mrk\,573, and NGC\,2992.  A minimum S/N cutoff was set such that spaxels with a low S/N were not inputted into the binning algorithm; the cutoff was ${\rm (S/N)_{min}} = 5$ for ESO\,138-G01 and ${\rm (S/N)_{min}} = 8$ for the other three galaxies.  Minor manual cleanup of the tessellation maps was required, which involved removing isolated spaxels and regularizing bin shapes in low-S/N regions.

The line fitting followed the approach used by \cite{Thomas_2017_S7DR2}.  Gaussian emission-line fitting was performed  using {\scshape lzifu} \citep{Ho_2016_LZIFU} to obtain the measured emission-line fluxes .  The {\scshape lzifu} code wraps {\scshape ppxf} \citep[][used to fit the stellar continuum for subtraction]{Cappellari_Emsellem_2004_ppxf} and {\scshape mpfit} \citep[][used to fit Gaussians to the emission lines]{Markwardt_2009_MPFIT}.  The fitting was run with 1-, 2- and 3-component fits for each spatial bin in each galaxy.  The velocity and velocity width for each component were fixed by {\scshape lzifu} for all fitted lines in a spectrum.  A small number of fitting failures occurred; at least one from a failure to fit the stellar continuum and other failures presumably due to the complex line profiles in the relevant spectra.  The ${\sim}2 - 6$ missing bins in each galaxy do not affect our conclusions.

An artificial neural network {\scshape lzcomp} \citep{Hampton_2017_LZCOMP} was used to select the optimal number of kinematic components from 1-, 2- and 3-component fits for each bin.  The neural network was trained on the same data as in the S7 Data Release 2 products \citep{Thomas_2017_S7DR2}.  The training data was produced by multiple astronomers selecting between 1-, 2- or 3- component fits to S7 cubes.  The final line fluxes used in the NebulaBayes analysis were the sum of the fluxes in the individual components selected by {\scshape lzcomp}.

\vspace{0.2cm}
\section{Application of NebulaBayes} \label{subsec:NB_method}
This section describes the approach to constraining the physical parameters of interest in each galaxy, including the choices of emission lines, priors and NebulaBayes settings.

Care is necessary when selecting the emission lines to use in a NebulaBayes analysis.  The observed lines do not necessarily all arise in the same gas, and the equal weighting of lines in the analysis may result in a `dilution' of the constraining power if a significant number of low-information lines are included.  The Bayesian method allows the freedom to combine data from all available emission lines in the analysis, but a judicious selection is necessary nevertheless. 

For the ENLR analysis we chose to use the 11 emission lines [\ion{O}{2}]$\,\lambda 3726+29$, [\ion{Ne}{3}]$\,\lambda 3869$, \ion{He}{2}$\,\lambda 4686$, \Hb, \OIII$\,\lambda 5007$, \ion{He}{1}$\,\lambda 5876$, [\ion{O}{1}]$\,\lambda 6300$, \Ha, \NII$\,\lambda 6583$, \SII$\,\lambda 6716$, and \SII$\,\lambda 6731$.  This list contains the standard strong optical lines, including all of those used in the optical diagnostic diagrams of Figure~\ref{fig:BPTVO}.  The species \ion{He}{2} is produced by photons with energies above the ionization potential of 24.6\,eV and hence its line fluxes are sensitive to the hardness of the ionizing AGN continuum.

For the \HII\ region analysis we chose \OII$\,\lambda\,3726 + 29$, \OIII$\,\lambda\,4363$, \OIII$\,\lambda\,5007$, \ion{He}{1}$\,\lambda\,5876$, \ion{O}{1}$\,\lambda\,6300$, \NII$\,\lambda\,6583$, and \SII$\,\lambda\,6716,6731$.  We also include \Ha\ and \Hb\, which are essential for normalization and reddening-correction.

The equal weighting of emission lines in the NebulaBayes analysis does not take advantage of our prior knowledge of physical parameters revealed by particular diagnostic line ratios.  For simplicity we apply only one prior in general, on the flux ratio \SII$\,\lambda\,6716$ / \SII$\,\lambda\,6731$, which is an electron density diagnostic so constrains the gas pressure.  We required an additional contribution to the prior for ESO130-G01, described in Section~\ref{sec:ESO_special_treatment}.

We used traditional BPT/VO classifications to select only `Seyfert' Voronoi bins for the analysis.  The Seyfert classification was required on both the \NII\ and \SII\ diagrams (Figure~\ref{fig:BPTVO}); the [\ion{O}{1}] diagram provides little extra classifying power for these data, and was not used.  A bin was included in the ENLR analysis if and only if its line flux ratios lie above the \citet{Kewley_2001_starburst} theoretical extreme-starburst line on the \NIIHa\ diagnostic diagram and above both the \citet{Kewley_2001_starburst} theoretical extreme-starburst line and the \citet{Kewley_2006_AGN_hosts} empirical LINER/Seyfert separation line on the \SIIHa\ diagram.  We note that shock, \HII-region and diffuse ionized gas (DIG) emission may contaminate these spectra despite our cuts; this is further explored in the discussion of the results.

We took advantage of the ability of NebulaBayes to deredden the observed fluxes at every point in the interpolated parameter space.  The dereddening is based on matching theoretical Balmer decrements, and this approach allows fair comparisons with regions of the parameter space with predicted Balmer decrements that are higher than usual.

We chose a sampling density of 50 points along each of the four dimensions of the interpolated NLR grids, for a total of $6.25 \times 10^6$ points in the predicted flux grid for each emission line. For the 3D \HII\ region grids we chose a sampling density of 200 points along each of the three dimensions, for a total of $8 \times 10^6$ gridpoints.  The statistical and systematic uncertainties in the results presented here are in general larger than the spacing between the raw (photoionization model) gridpoints, so the exact sampling of the higher-frequency interpolated points is not expected to significantly influence the results.

The default fractional systematic grid uncertainty of $\epsilon = 0.35$ was used.  This parameter directly increases the variance $V_i$ in equations~\ref{eq:usual_likelihhod_contribution} and \ref{eq:UB_reduced}.

\vspace{0.2cm}
\section{Results} \label{sec:Results}
In Section~\ref{sec:nuclear_results} we present and discuss the results for the galactic nuclei.  In Section~\ref{sec:results_maps} we map the spatial distributions of parameter values in the `pure Seyfert' galaxies, and in Section~\ref{sec:param_results} we briefly describe the key features of the parameter maps for each galaxy.  In Section~\ref{sec:gal_results} we combine interpretations of the various measurements with prior knowledge of the galaxies to glean insights into the properties of each object.

\subsection{Nuclear spaxels} \label{sec:nuclear_results}
Below we present and discuss the constraints on the physical parameters obtained from NebulaBayes analyses of the nuclear spectra of each selected S7 galaxy.

\subsubsection{Competing solutions for ESO\,138-G01} \label{sec:ESO_special_treatment}
\begin{figure}
	\centering
	\includegraphics[width=0.47\textwidth]{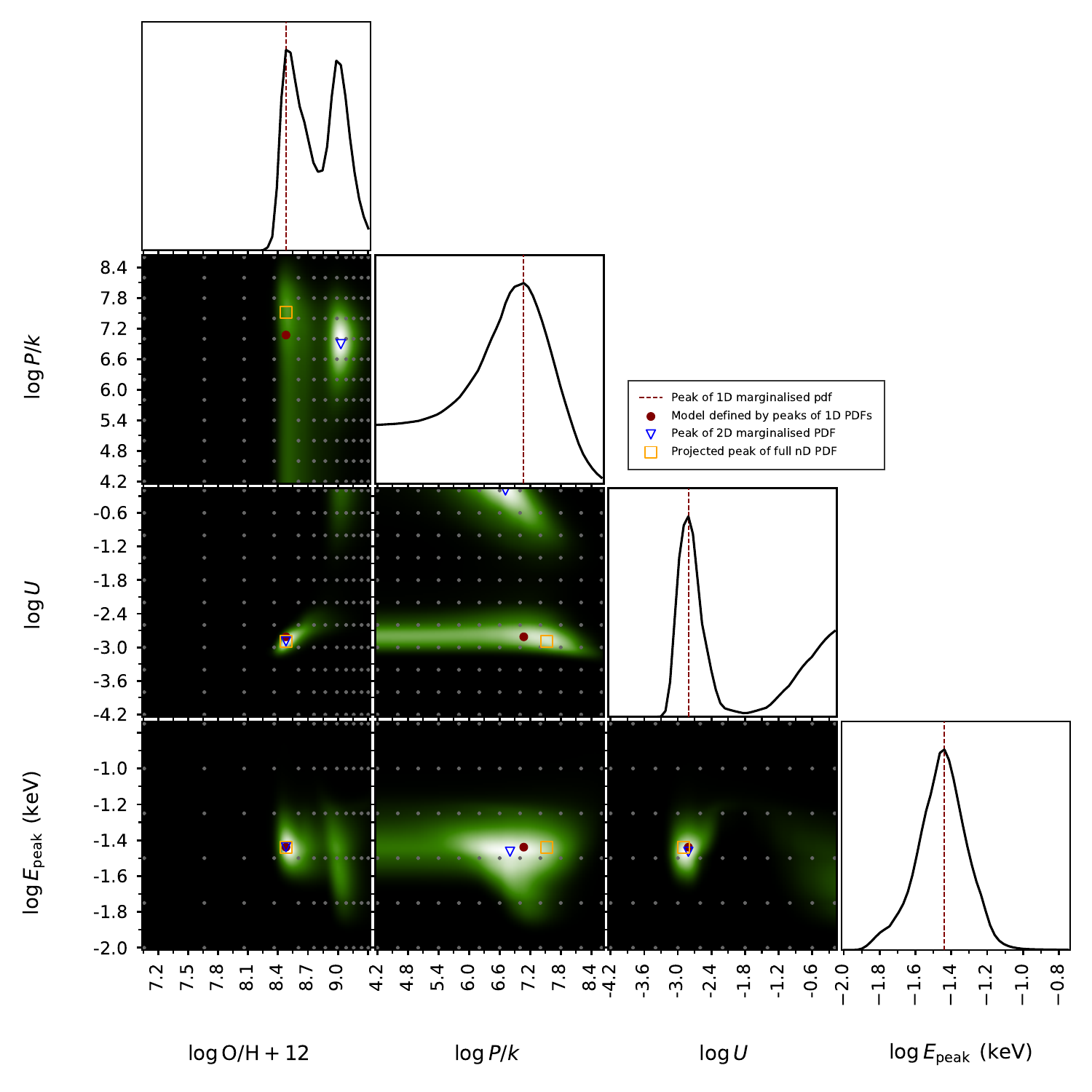}
	\caption{Likelihood for a nuclear bin in ESO\,138-G01.  There is a degenerate solution for $12 + \log$\,O/H and $\log U$, discussed in the text.  A prior was applied on the ionization parameter to select the lower-ionization parameter, lower-abundance solution, which is consistent with adjacent bins (Section\,\ref{sec:ESO_special_treatment}).  Parameter estimates are given in Table\,\ref{table:nuclear_results}.  \label{fig:ESO138nucLikelihood}}
\end{figure}

The nuclear region of ESO\,138-G01 required special treatment.  The likelihood produced by NebulaBayes for a nuclear spaxel is illustrated in Figure\,\ref{fig:ESO138nucLikelihood}, which shows two degenerate solutions.  The figure shows all possible combinations of marginalized 2D joint posterior PDFs, with the PDFs colored black to white through green from low to high probability density.  The gray points indicate the locations of the (uninterpolated) model gridpoints.  All four 1D marginalized PDFs are illustrated along the diagonal.

The ENLR gas in ESO\,138-G01 is stratified, with higher-excitation lines arising in higher-density environments closer to the nucleus \citep{Alloin_1992_ESO138G01}.  The photoionization models do not include dust destruction or cover the extreme regions of parameter space required to properly model the coronal lines, so these lines were not included in the NebulaBayes analysis.  Nevertheless, the two competing solutions evident in Figure\,\ref{fig:ESO138nucLikelihood} may reflect the different physical regions emitting the high- and low-ionization lines.

One solution, with an ionization parameter of $\log U \sim -3$ and an oxygen abundance of $12 + \log {\rm O/H} \sim 8.5$, appears to correspond to the normal NLR gas.  The other solution has an ionization parameter pushing the upper limit of the grid and much higher abundances, and may correspond to the coronal-line emitting gas, or alternatively may simply be a false solution.  If this high-$U$ solution is in fact caused by contamination of the strong emission lines by the coronal-line gas, the higher inferred abundances would presumably be due to the presence of species liberated from dust, which would provide a greatly enhanced cooling efficiency in the ionized plasma.

We selected the low-$U$ solution because it has reasonable parameter estimates that are consistent with adjacent spatial bins.  This solution was chosen by applying an extra prior that was a function of only $U$.  The extra prior was uniform (in logarithmic space) below $\log U = -2.0$, but exponentially decreasing above this value to exclude the high-$U$, high-metallicity solution.  We combined the prior on $U$ with the usual prior on the \SII\ ratio in the analysis of all bins in the galaxy, although the $U$-dependence only affected the results for the nuclear spaxels.

Although the nuclear metallicity measurements may be biased by contamination from the coronal line gas, the consistency between the nuclear results and the metallicity in adjacent, non-contaminated bins (Section~\ref{sec:results_maps}) suggests that the metallicity estimates have not been significantly affected.  Nevertheless, the upper uncertainties on the oxygen abundance are much larger in the nuclear region of ESO138-G01 than for the other galaxies (Table~\ref{table:nuclear_results}).

\subsubsection{Results for the Seyfert galaxies}
\begin{figure}
	\centering
	\includegraphics[width=0.47\textwidth]{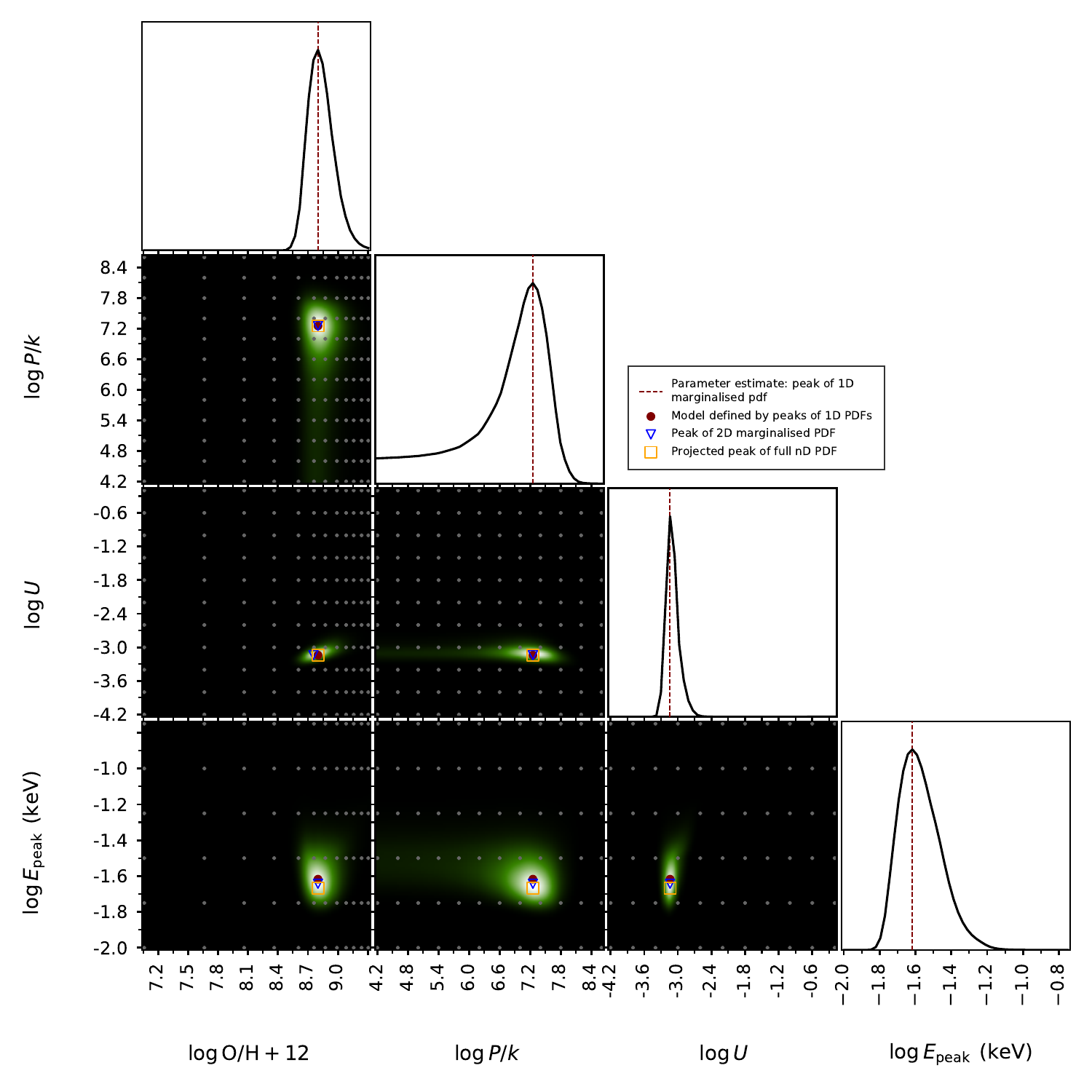}
	\caption{NebulaBayes posterior for a nuclear bin in IC\,5063.  Parameter estimates are given in Table\,\ref{table:nuclear_results}.  \label{fig:IC_5063nucPosterior}}
\end{figure}

\begin{figure}
	\centering
	\includegraphics[width=0.47\textwidth]{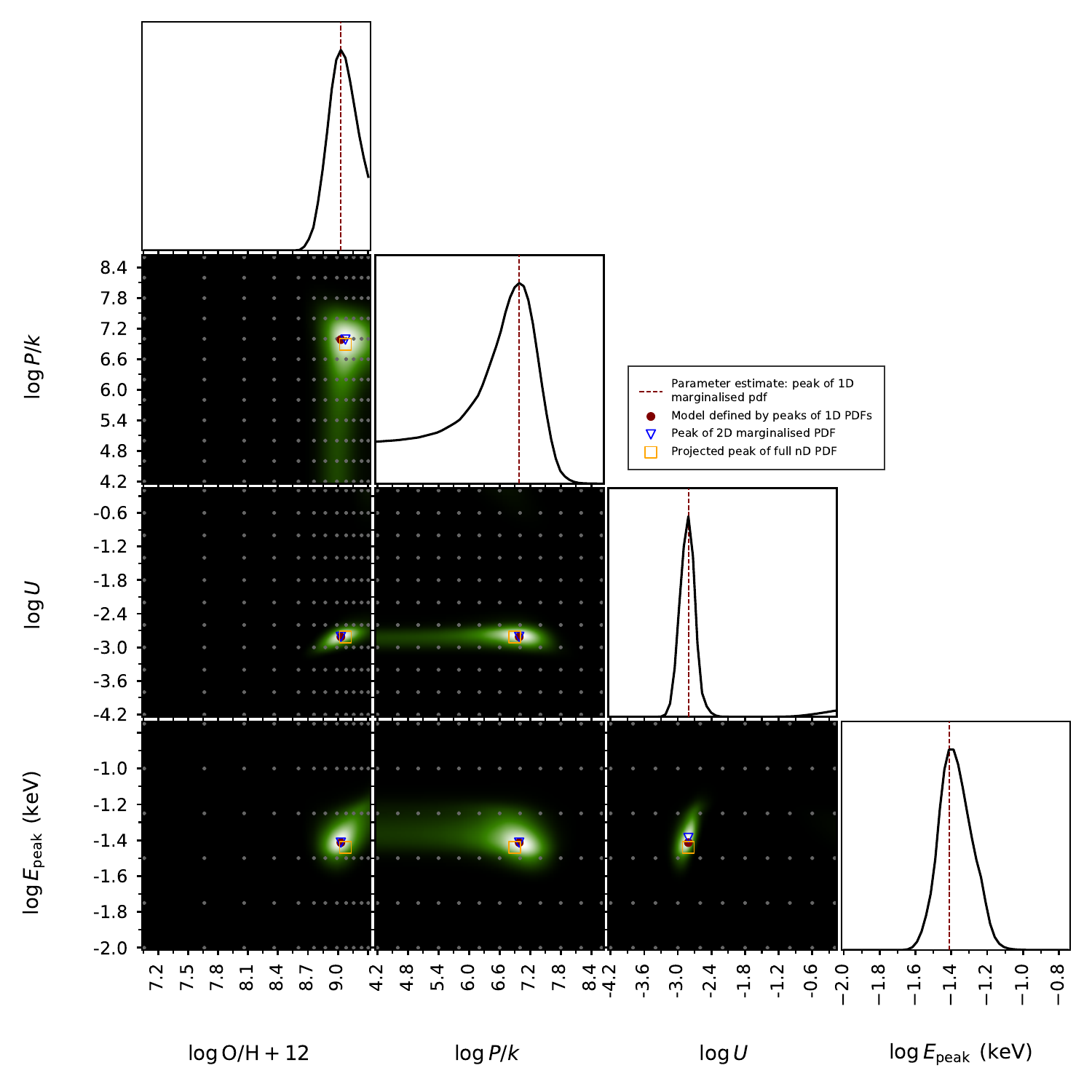}
	\caption{NebulaBayes posterior for a nuclear bin in Mrk\,573.  Parameter estimates are given in Table\,\ref{table:nuclear_results}.  \label{fig:Mrk_573nucPosterior}}
\end{figure}

\begin{figure}
	\centering
	\includegraphics[width=0.47\textwidth]{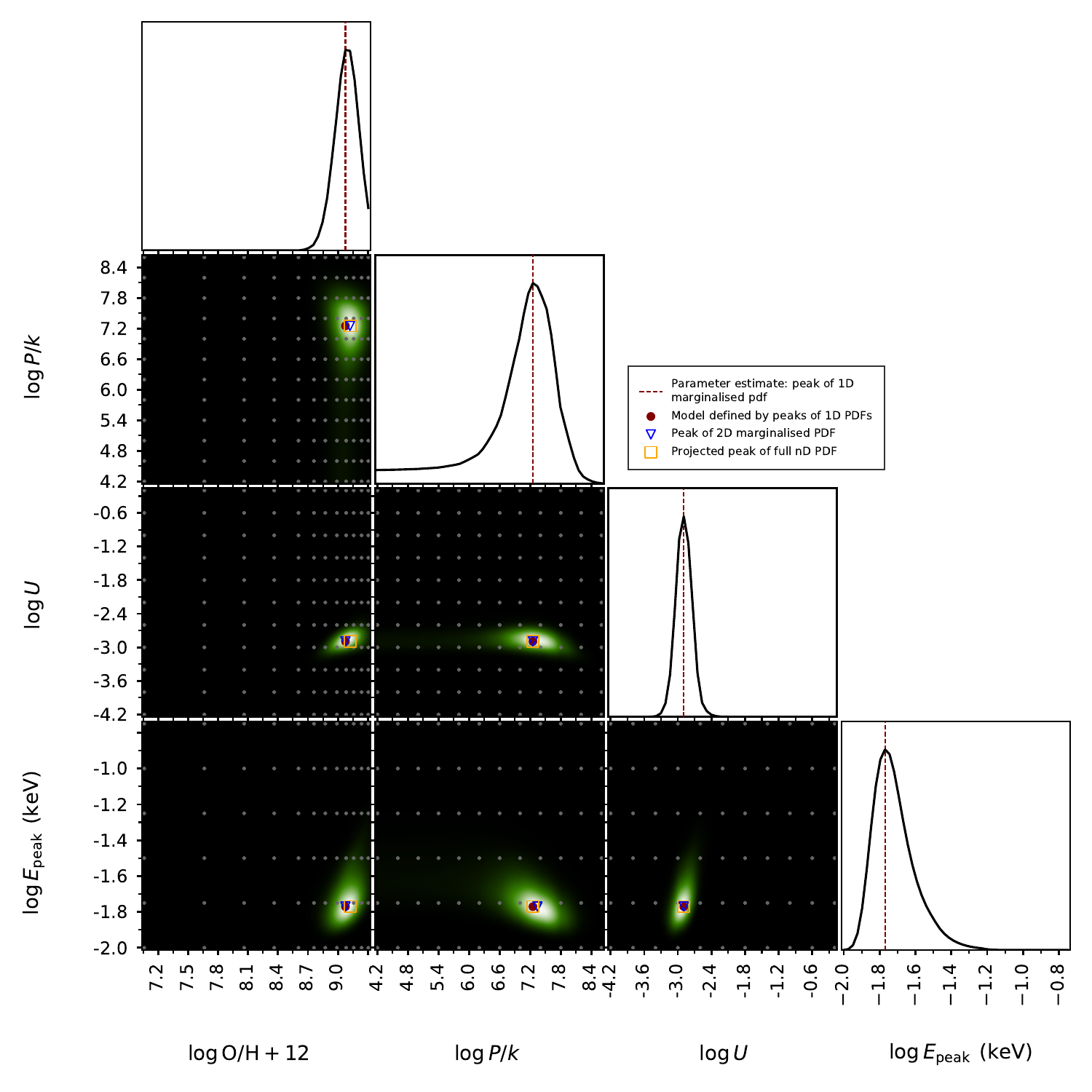}
	\caption{NebulaBayes posterior for a nuclear bin in NGC\,2992.  Parameter estimates are given in Table\,\ref{table:nuclear_results}.  \label{fig:NGC_2992nucPosterior}}
\end{figure}

Posteriors for the nuclear bins for the other three selected Seyfert galaxies are shown in Figure\,\ref{fig:IC_5063nucPosterior} for IC\,5063, Figure\,\ref{fig:Mrk_573nucPosterior} for Mrk\,573, and Figure\,\ref{fig:NGC_2992nucPosterior} for NGC\,2992.

The NebulaBayes parameter estimates for all four galaxies are presented in Table~\ref{table:nuclear_results}.  The table shows that measured nuclear values of \Ep\ vary from $\log E_{\rm peak} \mathrm{(keV)} = -1.77$ (\Ep$ = 17$\,eV) in NGC\,2992 to $\log E_{\rm peak} \mathrm{(keV)} = -1.41$ (\Ep$ = 39$\,eV) in Mrk\,573.  The nuclear abundances are half Solar in ESO\,138-G01, 1.2~Solar in IC\,5063 and approximately twice Solar in Mrk\,573 and NGC\,2992.  This variation in nuclear abundances is surprising, considering the small range in stellar masses, and we discuss the anomalously low values for ESO\,138-G01 and IC\,5063 in Section~\ref{sec:gal_results}.

\begin{DIFnomarkup}
	\begin{deluxetable*}{lcccc}
		\centering
		\tabletypesize{\scriptsize}
		\tablewidth{450pt}
		\tablecaption{NebulaBayes parameter estimates (peak in 1D marginalized posterior PDFs) for a nuclear spaxel in each `pure Seyfert' galaxy, with bounds of the 68\% credible intervals. \label{table:nuclear_results}}
		\tablehead{
			\multicolumn{1}{l}{Galaxy} & \multicolumn{1}{c}{$\log \, E_\mathrm{peak}\;\,\mathrm{(keV)}$} & \multicolumn{1}{c}{$12 + \log \, \mathrm{O/H}$} & \multicolumn{1}{c}{$\log \, P/k\;\,({\rm cm}^{-3}\,{\rm K})$} & \multicolumn{1}{c}{$\log \, U$}
		}
		\startdata
		ESO138-G01 &  -1.44$^{+0.18}_{-0.08}$ &  8.48$^{+0.46}_{-0.05}$ &   7.43$^{+0.36}_{-2.0}$ &  -2.81$^{+0.41}_{-0.16}$ \\
		IC5063     &  -1.64$^{+0.18}_{-0.08}$ &  8.85$^{+0.18}_{-0.14}$ &   7.16$^{+0.27}_{-1.3}$ &  -3.06$^{+0.16}_{-0.16}$ \\
		MARK573    &  -1.41$^{+0.15}_{-0.08}$ &  9.08$^{+0.14}_{-0.18}$ &   6.98$^{+0.18}_{-1.8}$ &  -2.73$^{+0.16}_{-0.24}$ \\
		NGC2992    &  -1.77$^{+0.18}_{-0.08}$ &  9.08$^{+0.14}_{-0.14}$ &  7.25$^{+0.36}_{-0.99}$ &  -2.89$^{+0.16}_{-0.16}$ \\[0.2cm]
		\enddata
	\end{deluxetable*}
\end{DIFnomarkup}

In Table~\ref{table:Z_comparison} we compare our derived metallicities to the estimates recently reported by \citet{Dors_2015_AGN_Z, Dors_2017_NLR_N_O} and \citet{Castro_2017_AGNZ}.  Our inferred metallicities are higher for IC\,5063 and Mrk\,573 than the comparison values, with the \citet{Castro_2017_AGNZ} estimates, \citet{Dors_2017_NLR_N_O} estimates, and all but one of the \citet{Dors_2015_AGN_Z} estimates lying below our 68\% credible intervals.  \citet{Dors_2017_NLR_N_O} allowed N/O to vary, which may have introduced systematic effects that contributed to the observed offsets.  However, there may be many other reasons for the apparent discrepancies between our measurements and the literature values, including aperture effects, systematic differences due to the use of different sets of emission lines, differences in the shape of the photoionizing spectrum, and systematic differences between models, including the photoionization codes themselves (\citet{Castro_2017_AGNZ} and \citet{Dors_2017_NLR_N_O} use CLOUDY).  One difference is that we use isobaric (constant gas pressure) models and include pressure as a parameter in our grid, whereas \citet{Castro_2017_AGNZ} use a fixed density of $n_e = 500$\,cm$^{-3}$.

\begin{DIFnomarkup}
	\begin{deluxetable}{lcccc}
		\centering
		\tabletypesize{\scriptsize}
		\tablewidth{15cm}
		\tablecaption{Comparison of Seyfert galaxy nuclear metallicity measurements\tablenotemark{1} with \citet{Dors_2015_AGN_Z,Dors_2017_NLR_N_O} and \citet{Castro_2017_AGNZ}\\[-0.4cm] \label{table:Z_comparison}}
		\tablehead{
			Galaxy &
			This work &
			Dors+15\tablenotemark{2} &
			Dors+17 &
			Castro+17    }
		\startdata
		ESO\,138-G01 & $0.5_{-0.1}^{+1.0}$ & 0.4, 0.6 & 0.3 & 0.6 \\
		IC\,5063     & $1.2_{-0.3}^{+0.6}$ & 0.6, 0.8 & 0.6 & 0.9 \\
		Mrk\,573     & $2.1_{-0.7}^{+0.8}$ & 0.8, 1.5 & 1.0 & 1.2 \\
		NGC\,2992    & $2.1_{-0.6}^{+0.8}$ & -        & -   & -   \\[0.1cm]
		\enddata
		\tablenotetext{1}{Metallicities reported as a fraction of Solar, assuming a Solar oxygen abundance of $12 + \log {\rm O/H} = 8.76$ where necessary.}
		\tablenotetext{2}{Estimates using two different calibrations from \citet{Storchi-Bergmann_1998_AGN_Z}}
	\end{deluxetable}
\end{DIFnomarkup}

A constant density does not allow for the important effects of radiation pressure on the structure of the NLR nebulae \citep[e.g.][]{Dopita_2002_Prad, Groves_2004_dusty_models}, and we show in recent work that varying the density is necessary to accurately model \HII\ region nebulae (Kewley et al. (2018), submitted). Our measured nuclear pressures of $\log P/k$\,(cm$^{-3}$\,K)~$ = 7.0 - 7.4$ correspond to $n_e \sim 500 - 1100$\,cm$^{-3}$ for a plasma with a temperature of $10^4$\,K.

\citet{Bennert_2006a_Sy2s_slit} studied a sample of six local Seyfert~2 galaxies with long-slit data, including IC\,5063.  The nuclear electron density of $n_e \sim 640$\,cm$^{-3}$ and temperature of 13900~K measured by the authors (their Table~5 and Figure~6) correspond to a pressure of $\log P/k$\,(cm$^{-3}$\,K)~$ = 7.3$ for a fully ionized plasma, which is consistent with our result $\log P/k$\,(cm$^{-3}$\,K)~$ = 7.16^{+0.27}_{-1.3}$ (Table~\ref{table:nuclear_results}).  \citet{Bennert_2006a_Sy2s_slit} estimated the nuclear ionization parameter using the \OII/\OIII\ ratio; their value of $\log U = -2.7$ compares to our measurement $\log U = -3.06^{+0.16}_{-0.16}$, with the difference presumably due to the significant differences in methodology.

\subsubsection{Results for the star-forming galaxy NGC\,4691}
\begin{figure}
	\centering
	\includegraphics[width=0.45\textwidth]{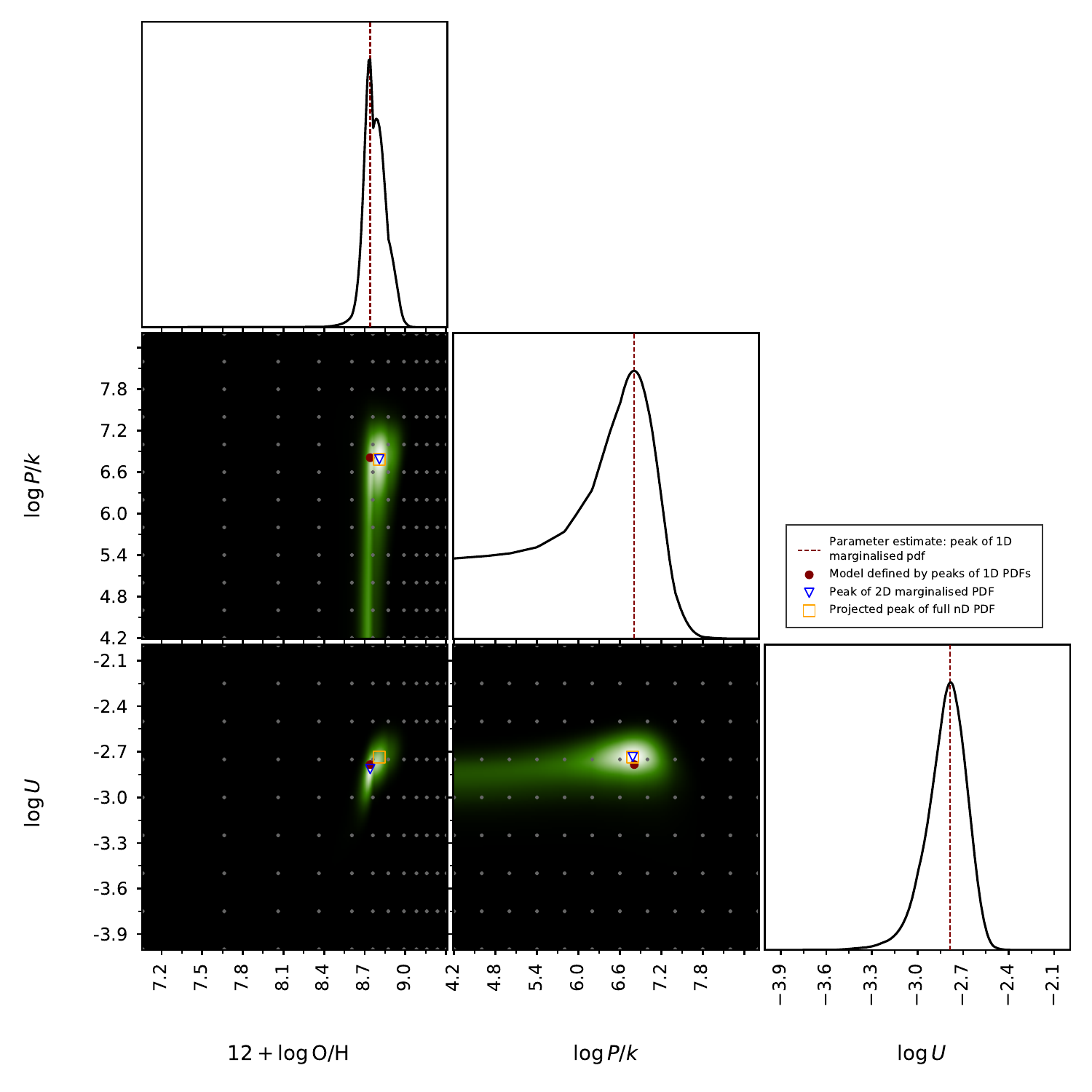}
	\caption{NebulaBayes posterior calculated using fluxes from an S7 nuclear spectrum for NGC\,4691, using the \HII\ region model grid.  The ionization parameter and metallicity are well constrained.  The pressure is not well constrained because the \SII\ ratio is near the low-density limit.  \label{fig:NGC4691_nuc_posterior}}
\end{figure}

To demonstrate the capability of NebulaBayes in constraining physical parameters for \HII\ regions, we include an analysis of the nuclear spectrum of the galaxy NGC\,4691.  This galaxy was included in the S7 sample due to a mis-classification in the source catalogue, and does not show any evidence for optical AGN activity.
The results of the analysis are shown in Figure\,\ref{fig:NGC4691_nuc_posterior}.

The parameter estimates (with 68\% credible bounds) are $12 + \log \mathrm{O/H} = 8.74_{-0.05}^{+0.12}$, $\log P/k$\,(cm$^{-3}$\,K)~$= 6.8_{-1.8}^{+0.2}$ and $\log U = -2.8_{-0.2}^{+0.1}$.  The approximately Solar nuclear metallicity is reasonable considering the estimated stellar mass of $10^{9.6}\,M_\odot$.  For comparison, the diagnostic of \citet{Dopita_2016_Zdiagnostic} gives $\log \mathrm{O/H} = 8.94$, which is 0.2\,dex higher.  The difference may be because the \citet{Dopita_2016_Zdiagnostic} diagnostic uses only four of the ten emission lines that we include in the NebulaBayes analysis, or perhaps because the diagnostic was calibrated to photoionization models based on an older version of MAPPINGS to the version used here.  Older atomic data was used, and \citet{Dopita_2016_Zdiagnostic} used ionizing stellar spectra directly from Starburst99 instead of from SLUG2.

\subsection{Maps} \label{sec:results_maps}
\begin{figure*}
	\centering
	\includegraphics[width=0.95\textwidth]{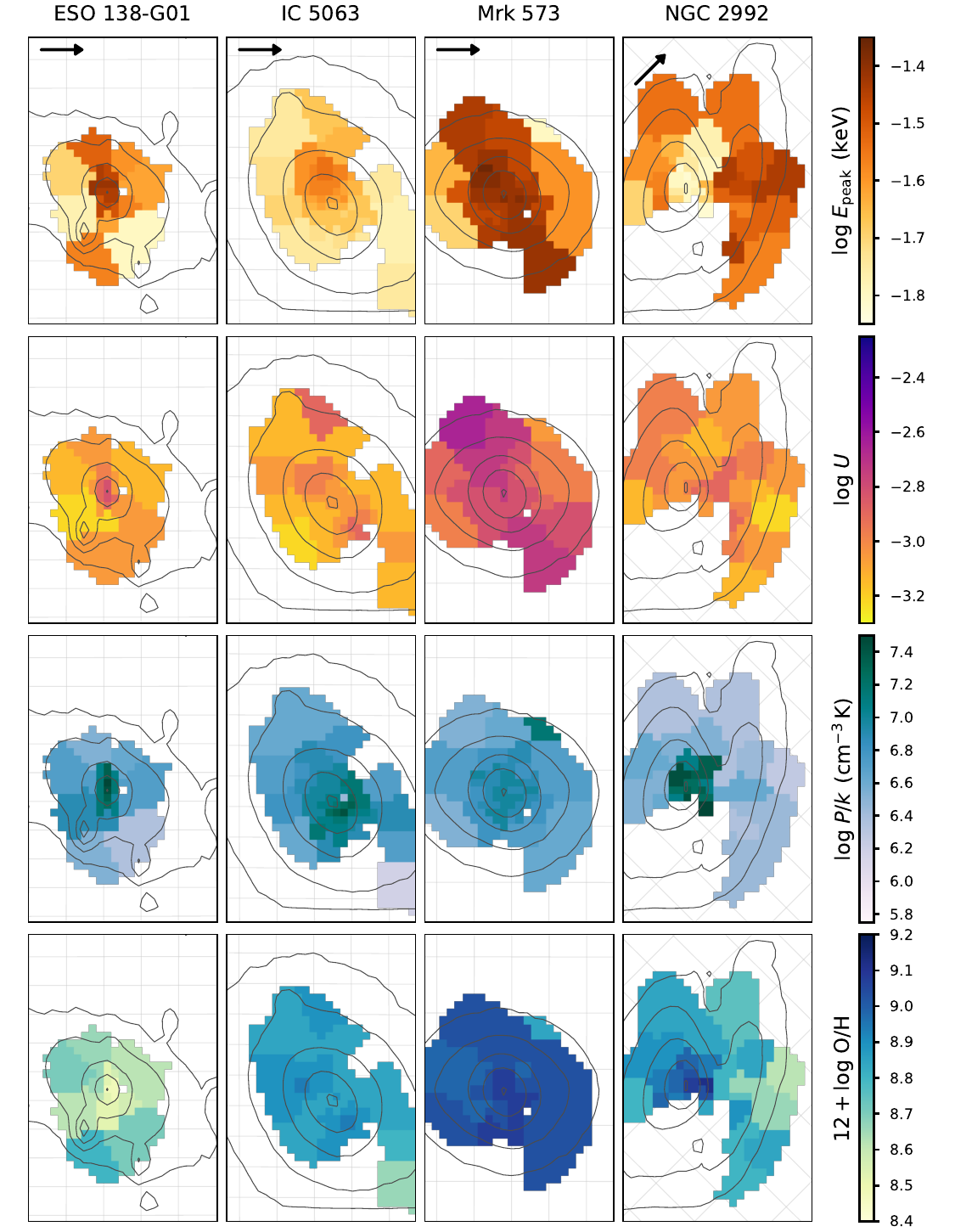}
	\caption{Maps of NebulaBayes parameter estimates for the four parameters for each of the four `pure Seyfert' galaxies.  Each measurement corresponds to the peak of the 1D marginalized posterior PDF for the relevant parameter.  The black arrows in the topmost panels indicate the direction of north and the logarithmically-spaced contours show the pseudo-continuum derived by summing the red S7 cube in the wavelength direction.  The WiFeS field of view is $25\arcsec \times 38\arcsec$, made up of $1\arcsec \times 1\arcsec$ spatial pixels.  Physical scales are shown in Figure~\ref{fig:Basic_maps}.  These results are discussed in Section\,\ref{sec:param_results} and Section\,\ref{sec:gal_results}.  \label{fig:Result_maps}}
\end{figure*}

\begin{figure*}[ht!]
	\centering
	\includegraphics[width=0.95\textwidth]{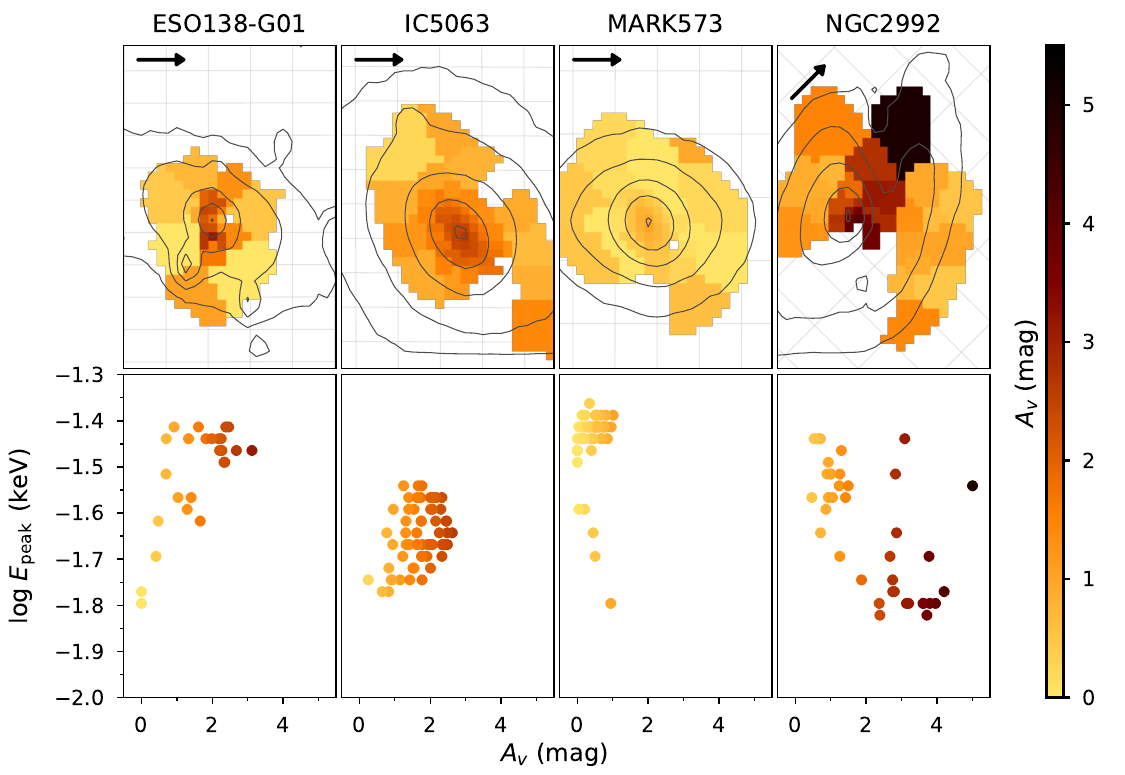}
	\caption{The distributions of inferred extinctions for each of the four `pure Seyfert' galaxies.  Top row: Maps of the inferred visual extinction $A_V$ (mag), similar to the maps in Figure\,\ref{fig:Result_maps}.  A dust lane is evident in the upper panel for NGC\,2992.  The dust lanes are not apparent in the panel for IC\,5063, likely because the observed nebular emission lies in the foreground.  Bottom row: \Ep\ versus the extinction $A_V$ for the same galaxies, with a point per spatial bin.  The `tails' of lower \Ep\ measurements for ESO\,138-G01 and Mrk\,573 are presumably due to spectral contamination.  There is no clear overall trend of \Ep\ with the degree of reddening, suggesting that local screening of the ionizing radiation by dusty gas is not significantly affecting the \Ep\ values (See Section~\ref{sec:more_discussion} for discussion). \label{fig:Av}}
\end{figure*}

Maps of parameter estimates for all four parameters across the ENLRs of all four `pure Seyfert' galaxies are presented in Figure~\ref{fig:Result_maps}.  Figure~\ref{fig:Av} contains similar maps of the inferred reddening, along with panels comparing \Ep\ to reddening.  The black arrows on the maps show the direction of north; east is 90$^\circ$ anticlockwise of north.  The physical scales are shown in Figure~\ref{fig:Basic_maps}; the width of the frames ranges from ${\sim}3$ to ${\sim}7$\,kpc.

\subsection{Notes on the parameter maps in Figure~\ref{fig:Result_maps}}  \label{sec:param_results}

\subsubsection{Hardness of the ionizing continuum}  \label{sec:Epeak}
Regions of relatively high inferred radiative hardness generally appear to coincide with the alignment of ionization cones in Figure~\ref{fig:Basic_maps}.  The possible causes of the measured \Ep\ variations are discussed in Section~\ref{sec:more_discussion}.  Screening of the ionizing radiation is expected to harden the spectrum, so the lower \Ep\ measurements outside the ionization cones are presumably due to spectral contamination rather than intrinsically lower \Ep\ values.

ESO\,138-G01:  The inferred \Ep\ values are highest in the center, but show a range of values in the larger outer bins.  The variation in values is larger than the uncertainties, which are $0.05 - 0.15$\,dex below and $0.13 - 0.20$\,dex above the estimated values.  The ENLR is evidently asymmetric, with harder radiation inferred to the west and east of the nucleus.  There is a `tail' of lower-\Ep\ values in some of the outer bins, which may be due to some combination of contamination from \HII\ regions or DIG and the lower reddening.

IC\,5063: \Ep\ values are systematically lower than in the other three `pure Seyferts'.  The smooth apparent decrease in \Ep\ in the central region, from southeast to northwest along the major axis, is comparable in magnitude to the 68\% uncertainty ranges of the individual bins ($0.05 - 0.1$\,dex below and $0.1 - 0.2$\,dex above the estimates).  The highest \Ep\ values are offset ${\sim}10$\,arcsec to the southeast of the nucleus.

Mrk\,573: The overall hardest inferred spectrum of the four Seyfert ENLRs studied here, with a large southeast-northwest swathe of the ENLR having \Ep$\, \sim 40$\,eV, and uncertainties of only ${\sim}0.1$\,dex below and ${\sim}0.15$\,dex above.  A small number of bins to the north and south of the nucleus show a softer \Ep\ value.  The `tail' of lower \Ep\ values may be due to star formation contamination, especially considering that the low-\Ep\ regions coincide with the outer ring/spiral features in Figure~\ref{fig:HST}.

NGC\,2992:  \Ep\ is measured to be higher in the ionization cones than in the central north-south dust lane.  The \Ep\ range of ${\sim}0.4$\,dex is larger than the uncertainties in individual bins, which are typically 0.1\,dex below and 0.15\,dex above.  \Ep\ gradients are evident in both the eastern and western ionization cones.  Overall the \Ep\ values are higher in the western cone, and the highest \Ep\ values occur over a range of radii in the western cone.

\subsubsection{Ionization parameter}  \label{sec:U}
The ionization parameter shows remarkably little variation both within or between galaxies, with all measured values falling within $-3.2 \leq \log U \leq -2.6$.  The widths of the 68\% credible intervals have a median of ${\sim}0.3 - 0.4$~dex for all four objects.

ESO\,138-G01:  The measured $U$ is somewhat higher in the nucleus, but the measurements only vary within 0.2~dex of the median $\log U = -3.1$, with a standard deviation of 0.1~dex.

IC\,5063:  Again the measured $U$ shows little variation.  Two points of interest are local regions of slightly elevated ionization parameter ${\sim} 5$~arcsec along the major axis either side of the nucleus.

Mrk\,573:  The highest ionization parameter measured in the four galaxies of $\log U = -2.65$ occurs in the southeastern part of the Mrk\,573 ENLR.  The distribution is asymmetric and bipolar; slightly higher $\log U$ values are observed along a southeast-northwest axis, with a slight positive radial gradient away from the nucleus in both directions along this axis.

NGC\,2992:  Both ionization cones and the dust lane show consistently similar $U$ values.

\subsubsection{Pressure}  \label{sec:Pressure}
All four ENLRs show the highest pressure in the nuclear region, with the pressure in the outer parts approximately a factor of 10 lower than the maximum.

ESO\,138-G01:  The highest pressures are measured in the narrow east-west nuclear region along the ionization cone axis, where a prior was required to choose the desired solution.  In the outer regions the pressures appear higher perpendicular to the cone axis compared to the on-axis pressures.

IC\,5063: The highest measured pressure is slightly offset from the nucleus to the northwest.

Mrk\,573: The nuclear pressure is lower than in the other galaxies (Table~\ref{table:nuclear_results}), and over the observed ENLR the pressures drop only 0.5\,dex (the smallest range of the four galaxies).  This apparent pressure homogeneity may reflect homogeneity of the entire ENLR (Section~\ref{sec:Mrk_573_results}).

NGC\,2992:  There is a remarkable discontinuity between the measured pressure in the dusty nuclear region (where 8 bins have $\log P/k$\,(cm$^{-3}$\,K)~$> 7.3$, with a maximum of 7.5), but lower pressures of $\log P/k$\,(cm$^{-3}$\,K)~$\leq 6.6$ in most of the remainder of the ENLR.  The ionization cones are presumably at a lower pressure because they are extraplanar.

\subsubsection{Abundances}  \label{sec:logOH}
Two of the four objects show a clear radial gradient in measured abundances.  Unfortunately there is no way to disentangle the geometry of the ENLR to measure de-projected gradients.

ESO\,138-G01:  There is a remarkable {\it inverse} abundance gradient, with the measured oxygen abundance ranging from approximately $12 + \log {\rm O/H} = 8.5$ (half-Solar metallicity) in the nucleus up to 8.8 in the outer regions.  Overall ESO\,138-G01 shows the lowest abundances of the four galaxies studied here.

IC\,5063:  Apart from an anomalous outer bin, this galaxy shows a tiny range in abundances, with $8.80 \leq 12 + \log {\rm O/H} \leq 8.94$.  There are two regions of slightly enhanced abundances either side of the nucleus, which have higher values than the nucleus.

Mrk\,573: Uniformly high abundances are observed across the ${\sim 6}$\,kpc of the ENLR, with the highest oxygen abundance of $12 + \log {\rm O/H} = 9.1$ (${\sim}3\,Z_\odot$) found in the nucleus, and measurements up to only 0.1\,dex lower elsewhere in the ENLR (except for another anomalous outer bin).

NGC\,2992:  A strong abundance gradient is observed in the ENLR gas, and lower abundances are found in the eastern ionization cone compared to the western cone.  Measured abundances decrease a massive 0.5\,dex between the maximum value measured in the nucleus of $12 + \log {\rm O/H} = 9.12$ and the 8.62 measured at the edge of the western ionization cone.

\subsection{Discussion of results for each galaxy} \label{sec:gal_results}

\subsubsection{ESO\,138-G01}  \label{sec:ESO138_results}
This is a disky, non-star forming galaxy that features luminous Seyfert~2 AGN activity.  The nuclear gas metallicity is slightly sub-Solar, despite the estimated stellar mass of $M_* = 10^{10.8}$\,$M_\odot$ \citep{Thomas_2017_S7DR2}.  In addition we measure an inverse abundance gradient in the ENLR, with abundances reaching up to ${\sim}1.1$ Solar in outer parts of the ENLR.  \citet{Alloin_1992_ESO138G01} found that this object harbors both an old stellar population and a younger population with an age less than 1\,Gyr.  

These observations could potentially be explained by a recent minor merger event.  In this scenario an infalling smaller galaxy would provide the lower metallicity gas, tidal effects in the merger would funnel the gas to the nucleus, and the relatively unobscured, lower-metallicity Seyfert AGN would result.  The radial inflow of accreted gas explains the inverse metallicity gradient, and star formation from the newly accreted gas would explain the younger stellar population.  This picture is consistent with the evidence that AGN activity generally lags star formation by a few hundred million years after a gas inflow  \citep[e.g.][]{Wild_2010_AGN_lag_SF}.

Despite the apparent recent arrival of gas into ESO\,138-G01, there does not appear to be evidence for current star formation. \citet{Goulding_2012_E138} fit a Spitzer-IRS spectrum to determine an AGN to starburst ratio of $\textrm{AGN:SB} > 0.9$, consistent with our optical emission-line diagnostics in Figure~\ref{fig:BPTVO}.

\subsubsection{IC\,5063}  \label{sec:IC_5063_results}
In this galaxy we measure all four parameters to be lower in the nucleus than in adjacent regions.  This is likely caused by the large-scale galactic wind, which is evident in the three Gaussian components required to fit the nuclear emission lines (with widths up to $\sigma_v \sim 350$\,km\,s$^{-1}$ and offsets between components of over 200\,km\,s$^{-1}$).  The observed nuclear ENLR gas appears to be dominated by outflowing foreground material -- the sum of the fluxes of the emission line kinematic components is dominated by the broad components in many nuclear bins.

The higher-$U$, higher-abundance `knots' either side of the nucleus on the major axis appear to avoid the SW-NE `stripe' of high velocity dispersion (Figure~\ref{fig:Basic_maps}).  \citet{Bennert_2006a_Sy2s_slit} also find that the ionization parameter does not peak in the nucleus, but is highest 5\farcs5 SE of the nucleus.  We speculate that these knots may be associated with lines of sight that avoid the illuminated outflowing extraplanar gas, allowing us to view gas that is more strongly illuminated, higher in abundance, and in fact closer to the nucleus than the gas observed in the nuclear bins.  Similarly this geometrical effect may explain the fact that the inferred pressure peaks immediately to the northwest of the nucleus.

The lower values in the nucleus than in adjacent regions are more difficult to explain for \Ep\ than for the other parameters, although the variation is similar in size to the uncertainties.  Because screening of the ionizing continuum is expected to increase rather than decrease \Ep\ (as described in Section~\ref{sec:more_discussion}), we instead favor shock contamination to explain the lower values.  We expect shocks to be associated with the high velocity dispersion in the outflow.  It is unclear how shock contamination may have affected the other parameter estimates.

The multi-phase outflow caused by the interaction between the radio jet and ISM in IC\,5063 is well studied \citep[][and references therein]{Dasyra_2015_IC_5063_SINFONI,Morganti_2015_IC5063, Congiu_2017_IC5063}.  \citet{Morganti_2015_IC5063} consider ALMA data and conclude that the rapidly outflowing $10^7\,M_\odot$ of molecular gas was most likely produced in post-shock cooling.  \citet{Dasyra_2015_IC_5063_SINFONI} use VLT SINFONI IFU observations to explore the complex 3D outflow structure in the central kiloparsec.  High velocity blue-shifted and red-shifted gas was found to be co-located with the north and south radio lobes, which demonstrates that acceleration and scattering of material by the radio jet powers the high velocity nuclear outflows.  \citet{Congiu_2017_IC5063} model line fluxes from long slit spectroscopy and argue that there is evidence for shocks significantly contributing to the ionization of high-velocity clouds.  The work of \citet{Morganti_2015_IC5063}, \citet{Dasyra_2015_IC_5063_SINFONI} and \citet{Congiu_2017_IC5063} lends support to our suspicion that shocks are contaminating the summed nuclear line fluxes.

We note that the S7 data for IC\,5063 exhibits \ion{Na}{1} D-line absorption in neutral outflows backlit by stellar light, which are being explored in detail by Rupke et al. (in preparation).

\subsubsection{Mrk\,573}  \label{sec:Mrk_573_results}
In this ENLR we see a double-cone of higher \Ep\ and $U$ values stretching from southeast to northwest across the field of view.  The ENLR features low reddening, with a median of $A_V = 0.4$\,mag increasing to only ${\sim}1$\,mag in the nucleus, despite uniformly high measured oxygen abundances of $1.7 - 2.1$~Solar.  The presence of coronal lines in the nucleus together with the low reddening and high measured abundances suggest that dust destruction is occurring in this object.  Dust destruction must have occurred in the nucleus to produce the coronal lines, but it may also be affecting the ENLR at large.  Destroying dust would lower the reddening and would also increase the oxygen abundances inferred from our dusty models.  This effect would occur because the dusty models need higher total abundances to achieve the same degree of cooling from free species as lower-depletion nebulae.

The N/O ratio provides further support to a picture in which large-scale dust destruction has occurred.  Across all 61 valid ENLR bins shown in Figure~\ref{fig:Result_maps} and Figure~\ref{fig:Av}, \OIII$\,\lambda 5007$ is underpredicted in all but two bins, with the median underprediction being 22\% of the dereddened observed flux.  Conversely \NII$\,\lambda 6583$ is consistently overpredicted, with a median overprediction of 9\% and overpredictions ranging up to 47\%.  The models used a depletion of 8.8\% of nitrogen and 22\% of oxygen by number onto dust grains.  There is more oxygen than nitrogen to be freed by dust destruction, so the underprediction of \OIII$\,\lambda 5007$ and overprediction of \NII$\,\lambda 6583$ is consistent with a scenario in which dust destruction has resulted in the N/O ratio being lower in the observed ENLR clouds than in the dusty models.

Confirming a scenario in which dust is destroyed across the ENLR is complicated by our lack of knowledge of the exact depletion of each element and the varying ratio of N to O abundances with metallicity.  A more careful analysis would be required to be certain that the N abundance is significantly lower than expected considering the O abundance.

The lack of a significant metallicity gradient over ${\sim 3}$\,kpc radially in the ENLR and the similar homogeneity of the measured pressures may indicate that radiation pressure-driven winds from the powerful AGN have mixed the ISM outwards.  We note that if ubiquitous dust destruction has caused a bad mismatch between physical and modeled depletions and gas-phase abundances, any intrinsic abundance gradient in the ENLR may be masked.  Nevertheless, if the AGN is responsible for both the lack of a metallicity gradient and large-scale dust destruction, the AGN may have been in a persistent high state for a considerable period of time.  The pervasive AGN influence may be causing an ongoing dearth of star formation through general disruption of the ISM -- destruction of dust shielding cool regions of the ISM, heating of the ISM, dissociation of molecular gas, and gas removal.

\citet{Fischer_2017_Mrk_573} studied gas kinematics in Mrk\,573 with the Gemini NIFS IFU and the Dual Imaging Spectrograph at Apache Point Observatory.  The authors concluded that outflowing gas currently extends to distances less than 1\,kpc, and invoked radiation pressure in the acceleration. 

\citet{Schlesinger_2009_Mrk573} used HST spectroscopy to study the outflow in Mrk\,573.  A combination of emission-line diagnostics, thermodynamics (measured pressures and energies) and other data led the authors to conclude that the ENLR nebulae are photoionized and that the AGN outflow is weak and likely thermal.  In particular the outflow was found to be unable to strongly shock material in the host galaxy.  The low \Ep\ measurements we observe in some outer bins are therefore unlikely to be associated with shock contamination; \HII\ region contamination may instead be the best explanation.

\subsubsection{NGC\,2992}  \label{sec:NGC_2992_results}
This galaxy exhibits a variety of different regions in the ENLR.  The high-extinction, high-pressure, low-\Ep\ nucleus is almost discontinuous with the eastern and western ionization cones.  The nuclear region features a broad Gaussian component in the emission line profiles with a velocity width often exceeding 600\,km\,s$^{-1}$.  The ionization cones show lower abundances, lower pressures, lower velocity dispersions and higher \Ep\ values than the nucleus.  The northernmost bin shows 5 mag of extinction, yet the measured parameters conform with the ionization cones rather than the highly-reddened nucleus.

The broad line profiles suggest the lower nuclear \Ep\ values may be attributable to spectral contamination by shocks (related to the AGN or perhaps the interaction with NGC\,2993), although \HII\ regions in the galactic disk may also make a contribution.  The softer ionizing continuum could also be due to a recent decrease in AGN activity, but the colocation of the broad outflowing lines and galactic disk with the low \Ep\ regions supports shock or \HII\ region contamination as the cause.  The varying \Ep\ in the ionization cones does not appear to be associated with variations in local dust obscuration, and may also be associated with shock or \HII\ region contamination.

The ENLR is most remarkable for the strong abundance gradient that is observed.  The maximum radial decrease is approximately 0.5\,dex in oxygen abundance over approximately $10\arcsec$ (1.5 projected kpc) in the western ionization cone in Figure\,\ref{fig:Result_maps}.  NGC\,2992 has a B-band $r_{25}$ of ${\sim}60\arcsec$ \citep{deVaucouleurs_1991_cat}, so the magnitude of the ENLR abundance gradient appears to be a large multiple of the typical gradients measured in star-forming disk galaxies, which are in the range -0.4 to -0.2\,dex\,$r_{25}^{-1}$ \citep{Sanchez_2014_CALIFA_Ograd, Ho_2015_Ogradients}.  The steep gradient may be due to the ionization cone being out of the plane of the disk, with the $z$-direction abundance gradient perpendicular to the disk being steeper than the radial gradient.

Unfortunately it is difficult to measure a robust deprojected and normalized gradient.  Any attempt at deprojection would need to take into account the unknown morphology of the ENLR and the high inclination of the galaxy \citep[$i \sim 70$\arcdeg;][]{deVaucouleurs_1991_cat}.  Tidal distortions and the high inclination both cause difficulties in accurately measuring $r_{25}$, so normalization of the gradient may also be challenging.  Nevertheless the absolute value of the abundance change is large and the physical gradient is surely relatively steep compared to typical radial gradients in disk galaxies.

Large-scale gas flows in this galaxy may be occurring due to the tidal interaction with its companion and AGN-driven winds; the large abundance gradient may be evidence that AGN outflows have been ineffective at redistributing metals and that tidal forces have so far been ineffective at funneling lower-metallicity material towards the nucleus.  Alternatively, tidally-driven low-metallicity gas out of the plane of the galaxy may have been illuminated by the AGN in the ionization cones.  A faint `bridge' of material between NGC\,2992 and its companion is visible in DSS images; the extraplanar low-metallicity gas may even originate in the companion galaxy.

\citet{Stoklasova_2009_IFU_N2992} use optical integral field data to study the central region of NGC\,2992.  A detailed kinematic analysis shows that the emitting species have different line-of-sight velocities, with the [\ion{O}{1}] and \SII\ line velocities varying most from the other strong lines (our {\scshape lzifu} fitting did not permit velocity differences between lines).  Stellar population modeling by \citet{Stoklasova_2009_IFU_N2992} reveals a population of young stars ${\sim}2$\arcsec\ NW of the nucleus.  This result lends some support to the idea of \HII\ region contamination in the nuclear spectra lowering our \Ep\ measurements, but shocks due to the AGN or tidal forces from the ongoing galactic interaction may be more significant contaminants.

High spatial resolution NIR integral field data from SINFONI is studied by \citet{Davies_2007_SINFONI_N2992}, who investigate nuclear star formation in Seyferts.  The data does not effectively constrain the star formation rate in the central $3 \times 3$\arcsec\ of NGC\,2992, although the authors find suggestions of recent star formation in the unresolved nucleus.

\cite{Davies_2016_rad_press} have studied NGC\,2992 using the same S7 data that we use in this work.  The authors visually compared the locus of flux ratios with a photoionization model grid on the BPT/VO diagrams (their Figure 3), inferring a range of ionization parameters of $-3.2 < \log U < 0$.  The fitting constraints in this work instead require the ionization parameter to be a relatively constant $\log U \sim -3.0$ and rule out values as high as $\log U \sim 0$ in the nucleus (Figure~\ref{fig:NGC_2992nucPosterior}; Table~\ref{table:nuclear_results}).  Our results suggest that the variation in line ratios on the BPT/VO diagrams is due to variations in the metallicity, ionizing continuum, and gas pressure rather than ionization parameter (Figure~\ref{fig:Result_maps}), although shock or \HII\ region contamination may also play a role, especially in the nuclear region.

\section{Further discussion} \label{sec:more_discussion}

We have presented the first robust, spatially-resolved measurements of abundances in AGN ENLRs.  The four `pure Seyfert' galaxies show very diverse distributions of metals in the ENLRs.  We measure a positive abundance gradient, a strong negative gradient, and nearly uniform high abundances across an ENLR.  Our results suggest that Seyfert AGN activity does not always flatten the gas-phase metallicity gradient in host galaxies by driving outflows of metal-rich nuclear material.  If this were the case we would not observe the inverse abundance gradient in ESO\,138-G01 or the strong positive abundance gradient in NGC\,2992.  However the large-scale, high-excitation, dust-poor, metal-rich, uniform-abundance ENLR in Mrk\,573 suggests that sustained AGN activity has the potential to regulate the ISM on kiloparsec scales.

An important insight is given by the contrast between a) the uniformity of the inferred ionization parameters and b) the ubiquitous radial decline in pressure in the ENLRs.  These results strongly imply that radiation pressure determines the density structure of the ISM on kpc scales.  The photon number density $n_{\rm ph}$ decreases with the inverse square of the radius; if the ionization parameter $U = n_{\rm ph} / n_{\rm H}$ is nearly constant, then the ISM density (proportional to $n_{\rm H}$) must decline radially at approximately the same rate.  This apparent similarity between the radial density distributions of photons and atoms occurs in all four ENLRs and must be caused by a coupling between the radiation field and the material illuminated in the ENLRs.  The only feasible coupling mechanism is radiation pressure.

We consider the radiation pressures predicted by the 1D MAPPINGS photoionization models for particular $\log U$ and $P$ values.  The $\log U$ measurements correspond to total radiation pressures $P_{\rm rad} / P_{\rm gas} \sim 2 - 10\%$, where $P_{\rm gas} \equiv P$ is the initial gas pressure.  This is a tight but ubiquitous range, which implies that $P_{\rm rad} / P_{\rm gas} \sim 2 - 10\%$ corresponds to a relatively widespread, long-lived and stable configuration of the ENLRs.  It is unclear whether this configuration arises only through $P_{\rm rad}$ influencing the internal structure of individual clouds, or if bulk radiation-pressure driven outflows of the ENLR material are also important.  Regardless of the exact effects, we can infer that $P_{\rm rad}$ is an important influence on the ENLR density structure even when $P_{\rm rad}$ is an order of magnitude weaker than $P_{\rm gas}$.

The evidence for an inverse-square density distribution is necessarily indirect.  We do not attempt to perform detailed 3D modeling of ENLR geometry or to deproject the observed pressure maps.  An inverse-square decrease in ISM density in the ENLRs is rather gradual considering that ENLRs are usually extraplanar.  We suggest that this relatively slow radial density decrease is evidence for outflows, including radiation pressure-induced outflows in the ionization cones.

In their long-slit study of six nearby Seyfert~2 galaxies, \citet{Bennert_2006a_Sy2s_slit} measured radial profiles in electron density and ionization parameter.  \citet{Bennert_2006b_Sy1s_slit} used similar methods to study a sample of six nearby Seyfert~1 galaxies.  The measured ionization parameters generally decreased with radius but were always within the tight range of $U \sim 1 - 5 \times 10^{-3}$ for the Seyfert~2 galaxies, with the Seyfert 1 galaxies often showing values above and below this range.

\citet{Bennert_2006a_Sy2s_slit, Bennert_2006b_Sy1s_slit} measured radial electron density gradients using the optical \SII\ doublet, finding power-law indices ranging from -1.3 to -0.8 for their Seyfert~2 galaxies and a range of -0.9 to -2.3 for the Seyfert~1 galaxies.  In many cases these measurements may be consistent with an inverse-square density distribution; difficulties with the procedure include projection effects, profiles that were generally asymmetric around the nucleus, and unknown temperature gradients causing systematic uncertainties in the density measurements.

Studying the interplay between the ISM density and radiation pressure is complicated by the fact that we observe clouds with a range of physical conditions and galactocentric radii along any given line of sight \citep[e.g.\ the locally optically emitting cloud (LOC) model of][]{Ferguson_1997_LOC_NLR_model}.  We note that the radiation pressure confinement (RPC) model of \citet{Stern_2014_NLR_RPC} predicts radial profiles of the ionization parameter and density in an ENLR; in the future this or similar models may be compared to the results of NebulaBayes analyses of S7 ENLRs.

We have also presented systematic measurements of the spatial variation in the ionizing AGN continuum across the observed ENLRs.  Unfortunately this parameter is almost certainly sensitive to contamination from \HII\ regions, LINER-like emission and the DIG, as well as screening effects due to gas and dust lying between the nucleus and the observed clouds.  Contaminating LINER-like emission is expected to be due to shocks, but may also arise from AGN radiation leaking out of the nuclear regions to excite nebulae with a low ionization parameter, and from excitation by post-AGB stars.  The likely contamination prevents us from drawing conclusions from the spatial variation in inferred \Ep\ values.  Almost all of the observed variation may be due to these confounding effects -- indeed our results may be consistent with all luminous Seyferts sharing a similar EUV spectrum with a peak at ${\sim}35 - 40$\,eV.  This scenario would be consistent with the speculation by \citet{Kraemer_1999_NLR_cont_abs} that NLR line ratio variations are affected more by absorbing material internal to NLRs than by intrinsic continuum shape variations.  Where variations in the incident continuum are not important, lower \Ep\ values would in fact {\it diagnose} contamination, which could reasonably be the case in IC\,5063 and NGC\,2992.

Screening by gas and dust is expected to harden the ionizing continuum, because the opacity of gas is highest at 1\,Ryd and decreases towards higher energies, and approximately the same is true for dust grains \citep[e.g.][]{Weingartner_Draine_2001_Dust}.  Figure~\ref{fig:Av} shows that inferred \Ep\ values do not appear to be strongly related to the degree of local dust extinction.  Screening associated with the low to moderate extinction may, however, have plausibly made a contribution to the small positive correlations between \Ep\ and $A_V$ for ESO\,138-G01 and IC\,5063, but contaminating \HII\ regions, DIG or shocks could also play a role.  Material that screens the ionizing continuum is not necessarily located in front of or even nearby to the gas excited by the hardened continuum.  Hence the absence of strong correlations in Figure~\ref{fig:Av} does not rule out absorption of the ionizing continuum causing the inferred spatial variations in \Ep.

Contamination by \HII\ regions is in some sense approximately accounted for by our \Ep\ parameter, which can take into account the resulting effective softening of the ionizing continuum (O stars have softer ionizing spectra than Seyfert nuclei).  Hence when \Ep\ is lowered by \HII\ region contamination, the metallicity measurement may remain reliable.  Evidence in the nuclei of IC\,5063 and NGC\,2992 suggests that shock contamination is also able to lower the inferred \Ep\ values (Section~\ref{sec:IC_5063_results} and Section~\ref{sec:NGC_2992_results}), however this effect is not as easily understood.  We do not consider the effects of contaminating DIG on the parameter estimates, but we note that DIG is likely to have a low surface brightness relative to NLR clouds, so DIG contamination is the least favored of the possible explanations for the measured \Ep\ variations.

Finally, we briefly consider how our results relate to variations in the available gas in the `pure Seyfert' galaxies.  These galaxies have no significant star formation (no `pure \HII' spectra), despite gas being available to feed the AGN.  The environment has likely contributed to fueling the AGN in NGC\,2992, which is interacting with a large companion galaxy, and in ESO\,138-G01, which has evidence for a recent inflow.  The observed large-scale AGN effects may have contributed to the lack of star formation, but conversely the extensive ENLRs may be associated with a lack of gas to hinder the ionizing radiation.  These galaxies feature old stellar populations and may have been caught in a short-lived phase when `quenching' is ongoing or recently completed \citep[e.g.][]{Leslie_2016_MS}.

\section{Conclusions} \label{sec:Conclusions}

We have presented the new code NebulaBayes, which implements a general method of comparing observed emission-line fluxes to photoionization model grids.  We use NebulaBayes with two grids -- a 4D NLR grid over metallicity, ionization parameter $U$, pressure $P$ and the hardness of the ionizing continuum \Ep, and also a 3D \HII\ region grid over metallicity, $U$ and $P$.  These grids are included with the code.

NebulaBayes was applied to four `pure Seyfert' galaxies from the S7 survey - ESO\,138-G01, IC\,5063, Mrk\,573 and NGC\,2992, as well as the nucleus of the star-forming galaxy NGC\,4691.

Our major conclusions are as follows:
\begin{enumerate} 
	\item We selected four Seyfert galaxies from the S7 survey with line emission dominated by ENLRs over kiloparsec scales.  These objects did not exhibit any \HII\ region-like spectra on optical diagnostic diagrams.
	\item We present robust, spatially resolved measurements of abundances in AGN extended narrow-line regions (ENLRs)
	\item We observe near-constant ionization parameters but steeply radially-decreasing pressures in all four ENLRs, implying that radiation pressure from the AGN has regulated the density structure of the ISM on kiloparsec scales
	\item We observe an {\it inverse} abundance gradient in the ENLR of the E-S0 galaxy ESO\,138-G01, and conclude that this galaxy experienced a recent inflow of relatively low-metallicity gas, possibly during a minor merger
	\item Our analysis of IC\,5063 is likely affected by contamination from shock excitation, and we find that the shock contamination seems to lower the inferred \Ep
	\item The ENLR of Mrk\,573 shows high values of \Ep\ and uniformly high metallicity.  There are indications that dust is substantially destroyed over kpc scales.
	\item NGC\,2992 shows steep metallicity gradients from the nucleus into the ionization cones, with the metallicity decreasing by up to half a dex.  This demonstrates that tidal forces from the companion galaxy and AGN outflows have not flattened intrinsic metallicity gradients.  Our analysis of the nucleus is likely contaminated by shock and possibly \HII\ region emission.
	\item We have presented systematic, spatially-resolved measurements of the hardness of the ionizing AGN continuum.  Contamination from shock excitation and possibly \HII\ regions has likely lowered the inferred \Ep\ values, so \Ep\ could potentially be used to diagnose contamination of NLR spectra
	\item We do not find convincing evidence that screening of the intrinsic ionizing continuum by gas and dust has resulted in harder measured values of \Ep.
\end{enumerate}

We anticipate that NebulaBayes will permit a greater understanding of how radiative AGN directly affect the ISM and hence star formation in galaxies.  In the future we may `mix' grids of NLR and \HII-region photoionization models to extend the work presented here to Seyferts in the most common type of hosts, star-forming late-type galaxies.  This would enable a systematic analysis of the Seyfert galaxies in the S7 sample.  The effects of AGN on radial abundance gradients will likely be both more important and more difficult to measure in the Seyfert-hosting late-type galaxies.

Our \HII\ region grid over the three parameters of metallicity, ionization parameter and pressure will enable more robust measurements of \HII\ region abundances in the future.

The NebulaBayes code is very general and could also be applied to other emission-line regions for which physical models are available, for example planetary nebulae.  In the future we may also use calculated grids of shock models with NebulaBayes to measure shock parameters in AGN-driven outflows.

\acknowledgments
We thank the anonymous referee for helpful comments.  M.D. and L.K. acknowledge support from ARC discovery project \#DP160103631.  Parts of this research were conducted by the Australian Research Council Centre of Excellence for All Sky Astrophysics in 3 Dimensions (ASTRO 3D), through project number CE170100013.  This research is supported by an Australian Government Research Training Program (RTP) Scholarship.  The authors made use of the National Aeronautics and Space Administration (NASA) Astrophysics Data System Bibliographic Services and the NASA/IPAC Extragalactic Database which is operated by the Jet Propulsion Laboratory, California Institute of Technology, under contract with NASA.

\software{{\sc mpfit} \citep{Markwardt_2009_MPFIT}, {\sc mappings\,v} \citep{Sutherland_Dopita_2017_shocks}, {\sc slug2} \citep{Krumholz_2015_SLUG2}, {\sc lzifu} \citep{Ho_2016_LZIFU}, {\sc ppxf} \citep{Cappellari_Emsellem_2004_ppxf}, {\sc lzcomp} \citep{Hampton_2017_LZCOMP}, {\sc numpy} \citep{Van_Der_Walt_2011_numpy}, {\sc NebulaBayes}}

\appendix

\section{Implementation of NebulaBayes} \label{sec:NB}
The NebulaBayes package is implemented in Python using the standard numerical library {\it numpy}  \citep{Van_Der_Walt_2011_numpy}, and works with both Python\,2 and Python\,3.  The code is available online\footnote{\url{https://github.com/ADThomas-astro/NebulaBayes}} and the package may be quickly installed from the Python Package Index (PyPI)\footnote{\url{https://pypi.python.org/pypi/NebulaBayes}\\ Install by typing: {\tt pip install NebulaBayes}}.

The code is distributed with the \HII\ region and NLR grids described in Section~\ref{sec:Grids}, however users may easily use their own custom grids.

The use of NebulaBayes consists of two steps:
\begin{enumerate}
	\item Initializing a NebulaBayes `model'. 
	\item Applying the initialized NebulaBayes model to observed emission-line data. 
\end{enumerate}
The (relatively computationally expensive) interpolation of predicted emission-line fluxes across the grid occurs during the first step.  The initialized model may then be applied relatively cheaply to a large number of sets of observed emission lines.

The inputs to the initialization step are the following:
\begin{enumerate}
	\item The input predicted model grid fluxes, specified in several possible ways, and associated parameter names
	\item Lists of emission-line names.  The model will only be initialized for these lines; hence the model can only later be applied to a subset of these lines.
	\item The number of sampling points along each dimension for interpolation of the emission-line grids
	\item The systematic relative error on grid fluxes, as a
	linear proportion (default is 0.35 or 35\%)
\end{enumerate}
The emission-line names are arbitrary and may refer to line blends or summed doublets as well as individual lines.  A low sampling density of interpolated points may be used during initial working and then a higher density used to obtain final results.  We note that the final parameter estimates are determined by the interpolated density -- a parameter estimate can effectively only take a value corresponding to one of the parameter values in the interpolated grid.

The key inputs to the model-running step are:
\begin{enumerate}
	\item The observed emission-line fluxes, errors, and line names.  Upper bounds are included by setting the flux to $-\infty$.
	\item A flag to indicate if NebulaBayes should perform a reddening correction (default False).  The code is able to de-redden the observed fluxes {\it at every gridpoint of the interpolated model grid} to ensure the best possible comparison between the model and observed fluxes.  This level of sophistication is necessary for NLR grids over which the variation of the intrinsic Balmer decrement is non-trivial, and is used for all the analysis in this work.  The de-reddening method is described in the appendix of \citet{Vogt_2013_HCGs_I}, and uses a relative extinction curve from \citet{Fischera_Dopita_2005_extinction}.  The line wavelengths must be supplied if de-reddening is requested.
	\item The prior.  There are several options available for customising the prior, and users may specify emission-line ratios to use as priors.  The default is a bounded uniform prior.
	\item A line to use for normalization.  NebulaBayes will normalize observed and model fluxes to the chosen line before making further calculations; \Hb\ was used in this work.
\end{enumerate}
Users are free to correct observed fluxes for reddening prior to using NebulaBayes.  The de-reddening method included in NebulaBayes is valid only over the wavelength range $2480 - 12390$\AA.

The optional prior is very flexible.  The default bounded uniform prior will be flat in linear or in logarithmic space for each parameter, depending on whether the grid values for the parameter are specified in linear or logarithmic space.  A user may supply a list of pairs of emission line names, in which case the NebulaBayes code will calculate a prior over the whole grid based on the ratio of each pair of lines (the density-sensitive optical \SII\ doublet, for example), using effectively the same formula as Equation~\ref{eq:naive_prob} (but comparing observed and predicted ratios, instead of single fluxes).  If multiple line-ratio priors are specified they will all be multiplied together (weighted equally).  This option allows users to treat known diagnostic line ratios as priors.  The line ratios are compared directly to the model grid, so these priors are similar to line-ratio diagnostics that are `calibrated on the fly' for the grid in use.  Line ratios are useful priors in NebulaBayes because the code otherwise treats the emission lines independently, whereas a specific line ratio (such as the N2O2 index) is able to provide `prior knowledge' (e.g. to constrain the metallicity).

When NebulaBayes is run on a set of observed lines, the following outputs are available:
\begin{enumerate}
	\item `Corner' diagrams containing a lower-triangular matrix of panels showing all possible joint 2D marginalized PDFs and 1D marginalized PDFs (as in Figures~\ref{fig:ESO138nucLikelihood} to \ref{fig:NGC4691_nuc_posterior}).  One of these figures may be written out for each of the likelihood, prior and posterior, and also for the contribution to the likelihood made by each individual line.
	\item A table of parameter estimates, including a row for each parameter and columns for lower and upper bounds on the 68\% and 95\% credible intervals.  Each parameter estimate is defined by the peak of the relevant 1D marginalized posterior PDF, and the bounds are calculated from fixed percentiles of the 1D PDF (e.g. 2.5\% and 97.5\% for the 95\% credible interval).
	\item A table comparing observed and (interpolated) modeled fluxes at the point of the parameter best estimates.
	\item In python code, NebulaBayes returns an object to allow programmatic manipulation of NebulaBayes data.  The object holds references to all of the data that NebulaBayes uses in calculations, including the likelihood, prior and posterior PDFs (as regularly-sampled $n$D arrays).
\end{enumerate}
NebulaBayes performs several checks to try to identify poorly-behaved/pathological posterior PDFs, the results of which are included in the output parameter estimate table.  These checks include testing if a parameter estimate is at the lower or upper limit of its modelled range, if more than 50\% of the probability is close to the lower or upper bounds, and if the parameter estimate lies within the credible intervals.

We remark on the following NebulaBayes implementation details:
\begin{enumerate}
	\item Emission-line fluxes are interpolated using linear interpolation, which is necessary to ensure consistent and reliable interpolation behavior over an arbitrary number of dimensions.  This method may lead to issues with interpolation accuracy, however it does guarantee that interpolated fluxes increase or decrease monotonically between raw input gridpoints, which is not the case for many standard interpolation routines.  Our results generally have uncertainties larger than the intervals between the model gridpoints (e.g. Figures~\ref{fig:IC_5063nucPosterior} to \ref{fig:NGC4691_nuc_posterior}), so the quality of the interpolation is not a significant issue in this work.
	\item The marginalization of PDFs uses trapezoidal integration, again because it is necessary to support an arbitrary number of dimensions.
\end{enumerate}
Users should consider if the density of gridpoints in a custom input photoionization model grid is sufficient for the desired precision of the NebulaBayes parameter estimates, taking into account the use of linear interpolation and trapezoidal integration.

NebulaBayes comes with a unit/regression test suite.  One of the tests uses fluxes from a gridpoint of a dummy model grid as the input observed fluxes, in order to ensure self-consistency of the NebulaBayes analysis -- i.e. that the output and input grid parameters match. 

Although they use very similar methods, there are some differences between NebulaBayes and IZI that we note here.  NebulaBayes is more general than IZI because it is agnostic to the set of parameters and works on more than two dimensions; NebulaBayes is supplied with a 3D \HII-region grid and a 4D NLR grid (Section~\ref{sec:Grids}) as opposed to the 2D \HII-region grid supplied with IZI (which does not have pressure as a parameter).  IZI uses spline interpolation, whereas NebulaBayes uses linear interpolation to handle an arbitrary number of dimensions.  Additionally, the treatment of upper bounds differs slightly between NebulaBayes and IZI, with the $P(f_i < e_i)$ in Equation~\ref{eq:UB_reduced} differing from the $P(f_i < 1 \sigma)$ approach used in IZI.

\end{document}